**Estimating the Impact of the Bitcoin Halving on Its Price Using Synthetic Control**

Minerva University

Vladislav Virtonen

Advisor: Prof. A. Diamond

April 2025

## Executive Summary

**Title:** Estimating the Impact of the Bitcoin Halving on Its Price Using Synthetic Control

**Name:** Vladislav Virtonen

**Background:** New Bitcoin is created through mining, where participants, known as miners, validate transactions and add them to the blockchain in exchange for newly issued Bitcoin. Bitcoin halving is a scheduled event that reduces the rate at which new Bitcoin is issued, cutting the reward for miners in half approximately every four years. This built-in supply reduction slows Bitcoin's inflation and reinforces its scarcity, making new supply increasingly limited. Historically, halvings have been associated with upward price trends, as lower supply and steady or growing demand can create upward pressure on the price. I have noticed many analysts and publications highlighting this association and even inferring causation when no one has conducted a proper causal analysis. However, correlation does not imply causation.

**Aim:** This is why I gathered data on the **2024 and 2020 Bitcoin halvings** and built an observational study using a causal model to control for other factors influencing Bitcoin's price (macroeconomic conditions, regulatory developments, growing adoption) and **isolate the effect of halving on price**. With this model, I intend to achieve a balance close to what a randomized control trial (RCT) would 'synthetically' because it is not feasible to conduct an RCT in this context – nobody can randomly assign cryptocurrencies to treated and control groups to conduct an interventional study like this. Only through achieving such balance is it possible to say that this specific treatment (halving) influenced the outcome (price) to a certain degree.

**Method:** I use **synthetic control**, which is a statistical method co-developed by Prof. Alexis Diamond, to estimate the causal effect of an event by constructing a synthetic version of the treated unit using a weighted combination of similar, unaffected units. Instead of comparing Bitcoin's price directly to another cryptocurrency (which may not be a good match), this method creates a synthetic Bitcoin from a mix of other cryptocurrencies that closely track Bitcoin's price movements before the halving. If the real Bitcoin price diverges from this synthetic version after the halving, I interpret that difference as the estimated impact of the halving. Synthetic control is ideal in my study because it helps create a more credible counterfactual – what Bitcoin's price might have looked like without the halving, while controlling for broader market trends.

**Findings:** I find evidence that the 2024 halving **positively affected** Bitcoin's price, with the estimated impact accounting for about one-fifth of Bitcoin's total percentage growth over the 17-month study period. However, when applying the same method to the 2020 halving, I do not find a statistically significant causal effect, likely due to broader market disruptions at the time when COVID-19 began. This is the first study to analyze Bitcoin halvings causally.

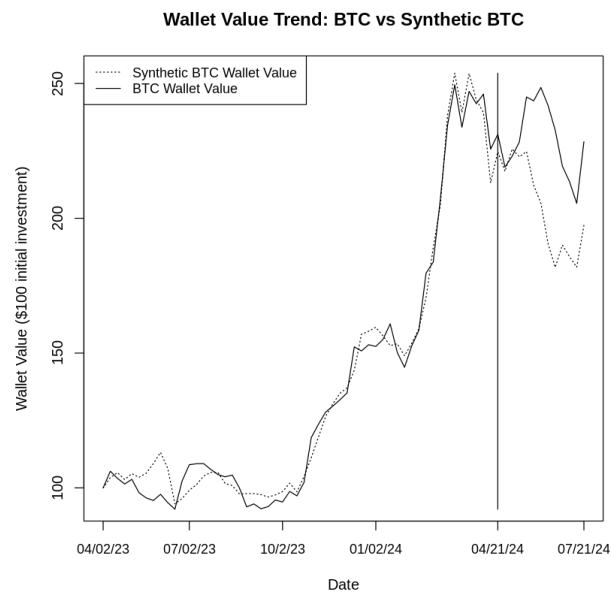

**Figure 3.** Wallet Value (a proxy of price) from April 2nd, 2023, until July 21st, 2024: Bitcoin versus synthetic Bitcoin. The vertical line shows the week of the 2024 Bitcoin halving. The divergence after the horizontal line is the estimated impact, calculated as the gap between the continuous and dotted lines.

**Estimating the Impact of the Bitcoin Halving on Its Price Using Synthetic Control**

**Abstract**

The third Bitcoin halving that took place in May 2020 cut down the mining reward from 12.5 to 6.25 BTC per block and thus slowed down the rate of issuance of new Bitcoins, making it more scarce. The fourth and most recent halving happened in April 2024, cutting the block reward further to 3.125 BTC. If the demand did not decrease simultaneously after these halvings, then the neoclassical economic theory posits that the price of Bitcoin should have increased due to the halving. But did it, in fact, increase for that reason, or is this a post hoc fallacy? This paper uses synthetic control to construct a weighted Bitcoin that is different from its counterpart in one aspect – it did not undergo halving. Comparing the price trajectory of the actual and the simulated Bitcoins, I find evidence of a positive effect of the 2024 Bitcoin halving on its price three months later. The magnitude of this effect is one-fifth of the total percentage change in the price of Bitcoin during the study period – from April 2nd, 2023, to July 21st, 2024 (17 months). The second part of the study fails to obtain a statistically significant and robust causal estimate of the effect of the 2020 Bitcoin halving on Bitcoin's price. This is the first paper analyzing the effect of halving causally, building on the existing body of correlational research.

**Introduction**

The phenomenon of Bitcoin halving has gained significant attention within both academic and non-academic worlds, particularly concerning its implications for Bitcoin's price stability and investment viability. Occurred on April 20th, 2024, and May 11th, 2020, the last two Bitcoin halvings represent a critical juncture for evaluating the largest cryptocurrency's economic dynamics and its purported role as a digital asset. Although prior research suggests Bitcoin halvings correlate with price increases (Meynkhard, 2019; Fantazzini & Kolodin, 2020), causal relationships remain unverified due to the challenge of observing counterfactuals.

This paper aims to bridge this gap by employing a synthetic control method (SCM) to construct a counterfactual version of Bitcoin – a "synthetic Bitcoin" – that does not undergo halving. This approach allows for estimating the halving event's causal effect on Bitcoin's price by comparing actual Bitcoin's price trajectory post-halving to that of the synthetic control. The methodology is predicated on the assumption that, by accurately modeling Bitcoin's price behavior in the absence of halving, one can isolate and quantify the specific impact of halving on its price dynamics. In other words, the paper presents a thought experiment for a counterfactual scenario in which the Bitcoin community reached a consensus on modifying the protocol to skip a scheduled halving event with a soft fork, similar to the famous Segwit User Activated Soft Fork (Pérez-Solà et al., 2019).

**Literature Review**

**Bitcoin**

Bitcoin, introduced in 2009, is the world's first cryptocurrency created, marking a significant departure from traditional, centralized monetary systems by employing a decentralized, peer-to-peer framework (Jin et al., 2022; Courtois et al., 2014). There is no consensus on the definition of Bitcoin's asset type in academic literature. Some view it as a security, a currency, or a commodity (Pagnotta & Buraschi, 2018). Bitcoin operates without a central bank, relying instead on a distributed network that facilitates transparent and secure transactions using cryptographic principles. The foundational technology of Bitcoin, blockchain, serves as a public ledger, recording all transactions and ensuring their veracity through a process known as "mining." This procedure demands that participants, or miners, employ significant computational resources to solve cryptographic hash puzzles, thereby validating transactions and adding them to the blockchain.

Despite its experimental inception and presentation as a novel form of electronic currency, Bitcoin's ascendance to mainstream financial markets has been meteoric, challenging conventional perceptions of what constitutes money (Courtois et al., 2014). Its value is influenced by factors other than just those affecting fiat (government-issued) currencies, such as inflation rates and governmental monetary policies. Some critical determinants include the network hashrate (a measure of the computational power of all participating nodes) and the cost of production, which reflect mining economics but have mixed impacts on price depending on market conditions (Ciaian, Rajcaniova, & Kancs, 2015; Fantazzini & Kolodin, 2020). Additionally, Bitcoin's volatility is influenced by macroeconomic variables, such as investor sentiment, transaction volumes, and speculative trading, as well as external correlations with traditional assets like gold during market stress (Taskinsoy, 2021; El Mahdy, 2024). Its volatility is particularly crucial in this research's panel data analysis.

The rapid adoption of Bitcoin by various financial institutions and governments, as well as its growing role as an investment vehicle, highlights the importance of better understanding the determinants of its price. Such an understanding is crucial for making informed investment decisions in the context of a currency that operates outside the traditional financial system (Jin et al., 2022).

**Halving**

Among digital currencies, Bitcoin stands out for its unique supply management approach. Bitcoin's halving mechanism, a feature of its deflationary design, enforces programmed scarcity by halving the mining reward, initially set at 50 BTC per block every 210,000 blocks (or approximately every four years), culminating in a finite supply of 21 million coins (Courtois et al., 2014). The process follows a geometric progression with a finite sum, ensuring that the total number of bitcoins never exceeds that final supply (El Mahdy, 2021). This system's deflationary model contrasts with traditional fiat currencies' inflationary tendencies controlled by central banks. The genesis of the halving mechanism was not explicitly outlined in Bitcoin's original whitepaper (Nakamoto, 2008); instead, it was introduced as a part of the software's design to control inflation by limiting the supply of new bitcoins entering circulation.

Due to the halving being widely anticipated, with fairly precise predictions of its timing, market participants' "animal spirits" may drive significant behavioral and price reactions prior to each halving event. Moreover, according to the Efficient Market Hypothesis, Bitcoin halving events should be priced in because they are predetermined and publicly known occurrences. In the context of synthetic control, this prompts the question of whether this scheduled event would even exert measurable effects (and whether we will observe 'preliminary' effects driven purely by the market's expectation of halving). I address these questions with in-time placebo tests in the 'Analysis' section.

| Cycle number | Year | Block Number | Block Reward | New BTC Mined | Total BTC Mined | % of all BTC mined |
|---|---|---|---|---|---|---|
| 1 | 2009 | 0 | 50 | 0 | 0.000 | 0.00000000% |
| 2 | 2012 | 210,000 | 25 | 10500000 | 10,500,000.000 | 50.00000000% |
| 3 | 2016 | 420,000 | 12.5 | 5250000 | 15,750,000.000 | 75.00000000% |
| 4 | 2020 | 630,000 | 6.25 | 2625000 | 18,375,000.000 | 87.50000000% |
| 5 | 2024 | 840,000 | 3.125 | 1312500 | 19,687,500.000 | 93.75000000% |
| 6 | 2028 | 1,050,000 | 1.5625 | 656250 | 20,343,750.000 | 96.87500000% |
| 7 | 2032 | 1,260,000 | 0.78125 | 328125 | 20,671,875.000 | 98.43750000% |
| 8 | 2036 | 1,470,000 | 0.390625 | 164062.5 | 20,835,937.500 | 99.21875000% |
| 9 | 2040 | 1,680,000 | 0.1953125 | 82031.25 | 20,917,968.750 | 99.60937500% |
| 10 | 2044 | 1,890,000 | 0.09765625 | 41015.625 | 20,958,984.375 | 99.80468750% |
| 11 | 2048 | 2,100,000 | 0.048828125 | 20507.8125 | 20,979,492.188 | 99.90234375% |
| 12 | 2052 | 2,310,000 | 0.0244140625 | 10253.90625 | 20,989,746.094 | 99.95117188% |
| 13 | 2056 | 2,520,000 | 0.01220703125 | 5126.953125 | 20,994,873.047 | 99.97558594% |
| | | | ... | | | |
| 33 | 2140 | 6,930,000 | 0.000000005820766091 | 0.004889443517 | 21,000,000.000 | 100.00000000% |

**Table 1**. Estimated timetable of new Bitcoin issuance.

**Neoclassical Economic Theory**

Schilling and Uhlig (2018) discuss the economic implications of competing currencies, including Bitcoin, by building an endowment economy model, highlighting how Bitcoin's predetermined supply growth affects its valuation and interacts with monetary policy aimed at fiat currencies. Meynkhard (2019) employs the Kendall rank correlation method to emphasize further the deflationary nature of halving events on Bitcoin's market value, discussing how reducing miners' rewards every four years enhances Bitcoin's scarcity and leads to a higher market price. This concept of scarcity is further elaborated by Radulović (2023), who synthesized previous research to conclude that Bitcoin's design—particularly the halving events—intentionally mimics the scarcity and value preservation characteristics of traditional deflationary assets. Bernardi and Bertelli (2021) propose a valuation framework for Bitcoin, similarly highlighting the 'value of scarcity' derived from its limited supply, akin to precious metals like gold, enabling Bitcoin's indefinite value growth. Krause, 2024 and Wu, 2024 point to the recent growing market demand for Bitcoin, stemming from increasing institutional adoption and financial product offerings like Exchange-Traded Funds (ETFs). Theoretically, as the supply of new Bitcoins entering the market decreases post-halving, the price should increase if demand does not drop.

While the existing literature suggests a correlation between Bitcoin's halving events and its price appreciation through the lens of neoclassical economic theory, history teaches us that economic theories do not always explain market outcomes accurately. Neoclassical economics, emphasizing equilibrium and rational expectations, can overlook cryptocurrency markets' complex nature and yield a prediction contrary to what can be observed empirically. Moreover, a significant portion of the existing analysis of Bitcoin halving commits the informal post hoc ergo propter hoc fallacy by suggesting that it has a positive 'effect' or 'impact' without addressing the issue of confounding. To truly understand the impact of Bitcoin halving on its price, adopting an empirical approach that contemplates the counterfactual and employs rigorous methods for estimating it is crucial.

**Methodology**

**Synthetic Control Model**

The synthetic control method (SCM) provides a robust approach for estimating causal effects in settings with a single treated unit and no natural comparison group. SCM is particularly advantageous in observational contexts where traditional comparison groups are unavailable and direct randomization is infeasible. SCM's utility in estimating the causal effect of Bitcoin halving lies in its ability to create a "synthetic Bitcoin" from other cryptocurrencies that did not undergo halving events, thus mimicking what Bitcoin's price trajectory might have looked like in the absence of halving. Such an approach is suitable for this study given that the event affects only one primary unit, Bitcoin, and there is no direct control group within cryptocurrency markets (the closest observed comparison groups are shown in Figures 1 and 2).

Originally developed by Abadie and Gardeazabal (2003) and later expanded by Abadie, Diamond, and Hainmueller (2010), SCM constructs a counterfactual for the treated unit—a "synthetic control"—by optimizing a weighted combination of untreated units to match the pre-treatment characteristics of the treated unit. There are J donor units. Unit j = 1 represents Bitcoin – the treated unit that experienced halving. Units j = 2 to J + 1 make up the donor pool. These units are potential donors for the synthetic control because they are not directly exposed to halving. Using the notation from Abadie et al. (2010), a synthetic control can be written as a Jx1 vector of weights $(w_1,..., w_{J+1})$ where $0 < w_n < 1$ and $w_2 +... w_{J+1}$ is equal to 1 (Abadie et al., 2014). In this study, only cryptocurrencies that do not undergo halving are included in the donor pool. The potential donors are shown in Appendix 3.

We choose the synthetic control, W, which minimizes the size of the difference between the pre-treatment intervention characteristics of the treated and the synthetic units. The difference is given by the vector $X_1 - X_0W$. Here, $X_1$ is the $k * 1$ vector of pre-intervention characteristics for Bitcoin and $X_0$ is the $k * J$ matrix of post-intervention characteristics for the control units (Abadie, 2021). This optimization process aims to guarantee the closest match between the synthetic control and the treated unit in terms of their pre-treatment characteristics. Once this optimization is completed, we treat the synthetic control unit as a counterfactual and compare its price trajectory post-halving with that of Bitcoin to evaluate the treatment effect.

Reporting the exact weights of the synthetic unit, the method ensures transparency. Also, the SCM avoids extrapolation (Abadie et al., 2010).

This paper does not aim to "prove" that the Bitcoin halving caused its price to increase. Instead, it aims to provide plausible evidence of impact under the SCM assumptions:

1. A linear factor model – a model where the outcome is determined by a set of observed and unobserved factors. The outcome is affected by a linear combination of unobserved common factors with time-varying effects. The SCM assumes that a weighted combination of control units can approximate the trajectory of the treated unit, conditional on observed and unobserved factors and that the idiosyncratic shocks are independent across units and time (Abadie, 2021).

2. Treatment isolation – the halving effect is only credible if no other interventions or shocks coincide with the treatment timing. This assumption is particularly questionable in the study of the effect of the 2020 Bitcoin halving on its price and is discussed further below.

3. SUTVA – the assumption of no interference between units, which is particularly bold in this context, as it assumes that the untreated cryptocurrencies are not indirectly influenced by Bitcoin's halving. This assumption is challenging in interconnected cryptocurrency markets, where Bitcoin's dominant position often drives broader market trends, creating a strong correlation between its price movements and those of other cryptocurrencies (Kulal, 2021). Bitcoin's halving events can influence investor sentiment and trading behavior across the entire market, potentially inducing spillover effects that may alter the trajectories of supposedly 'untreated' cryptocurrencies. These spillover effects can arise from various channels: investor rebalancing across assets, algorithmic trading responses to Bitcoin's price volatility, media-driven narratives that emphasize

Bitcoin's scarcity and potential value appreciation post-halving, etc. Despite this limitation, SCM offers a systematic and transparent framework that aligns well with the single-unit treatment structure and the goal of isolating Bitcoin's price response to halving.

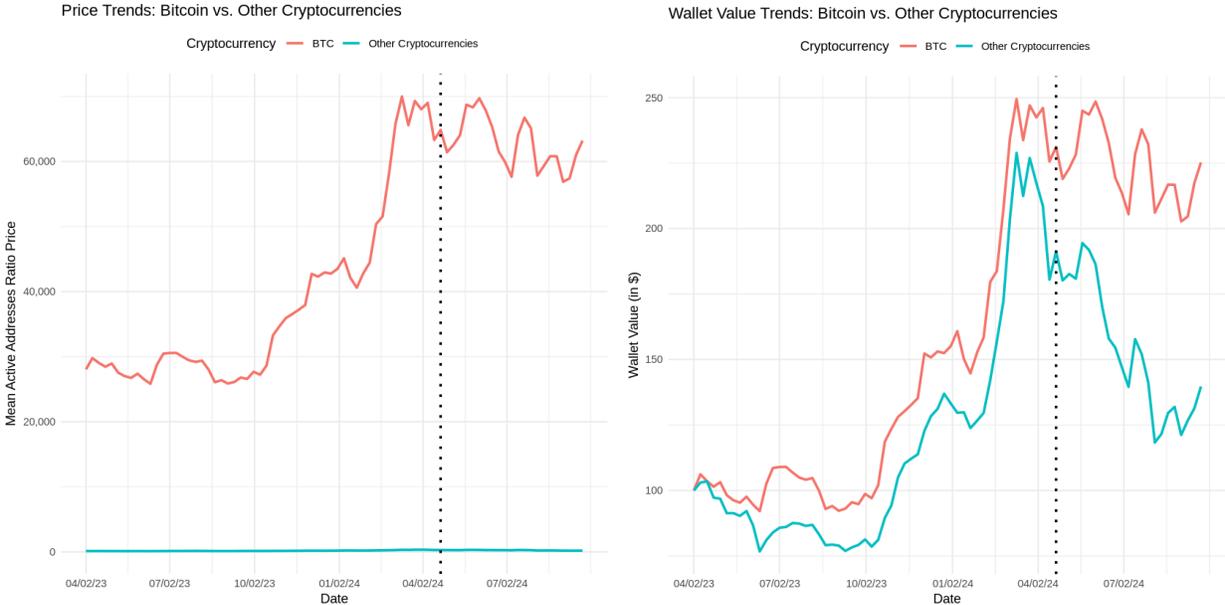

**Figure 1**. Price (on the left) and wallet value (on the right) for Bitcoin and other cryptocurrencies. The dotted line marks the 2024 Bitcoin halving. The blue line represents the average price or wallet value for the following cryptocurrencies (candidate donors): ADA, ALGO, ANKR, CRO, ENJ, ETH, FET, FTM, GNO, HOT, IOTX, KCS, LEO, LINK, LPT, MANA, MATIC, MKR, MX, NEXO, OKB, QNT, and TRX. The wallet value derivation is explained in the following section.

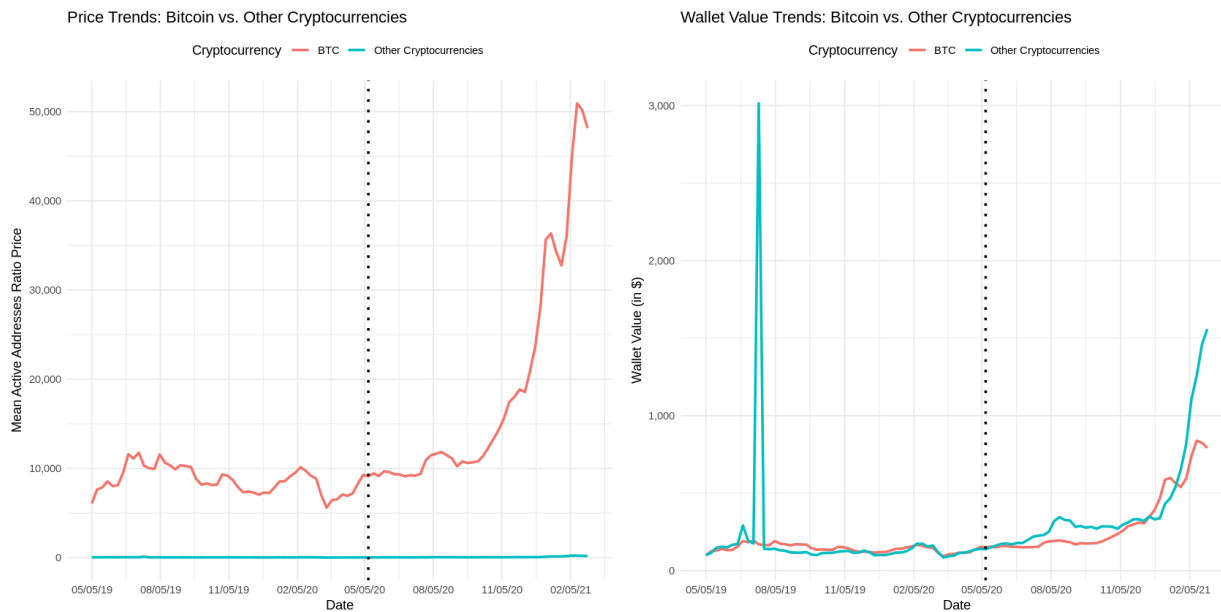

**Figure 2.** Price (on the left) and wallet value (on the right) for Bitcoin and other cryptocurrencies. The dotted line marks the 2020 Bitcoin halving. The blue line represents the average price or wallet value for the following cryptocurrencies (candidate donors): ADA, ANKR, CRO, DOGE, ENJ, ETH, FTM, GNO, HOT, IOTX, KCS, LINK, LPT, MANA, MATIC, MKR, NEXO, OKB, QNT, and TRX. The wallet value derivation is explained in the following section.

**Data**

I build panel weekly data using [IntoTheBlock](#)'s daily Market Intelligence financial, network, and ownership metrics for the variables in Table 2[1]. IntoTheBlock is a data science company providing real-time crypto market intelligence. The data for my sample comes from nodes of the top blockchains – mostly Ethereum.

The unit of analysis is cryptocurrency. I have weekly data for twenty-four cryptocurrencies, including the treated unit – Bitcoin. The time period for the 2024 halving ranges from April 2nd, 2023, to July 21st, 2024. The time period for the 2020 halving is from May 5th, 2019, to August 9, 2020. The treatment weeks are the weeks of the halvings – April 20,

---

[1] I outline the data sourcing and cleaning process in Appendix 1. As a result of the cleaning, I had to get rid of most variables, since they contained missing values. The original list of variables is shown in Appendix 2.

2024, and May 11, 2020, respectively.

My dependent variable in this study is Wallet Value (a proxy of Bitcoin price), which tracks the value of a $100 wallet investment in each cryptocurrency made in the week of April 2, 2023, in the model for the 2024 halving, and May 5th, 2019 in the model for the 2020 halving[2]. This approach standardizes the starting investment across cryptocurrencies, allowing a more direct comparison of the change in prices by accounting for the differences in cryptocurrency prices. Using wallet value instead of absolute price reduces the impact of initial price levels, as some cryptocurrencies in the dataset are valued under $1, while others exceed $50,000 (as shown on the left side in Figures 1 and 2). Another significant reason to transform the Y variable like this is that it is generally easier to construct a synthetic control when the treated unit's outcome (Y) lies within the range (or convex hull) of the donor units' outcomes. When the treated unit's Y values are within the range of the control units, the optimization problem is feasible, and the algorithm can find a convex combination of donor outcomes that closely replicates the treated unit's pre-treatment trajectory without extrapolating.

The initial quantity of each cryptocurrency in the wallet is determined by dividing $100 by the price on the baseline date. Subsequent weekly values reflect the changes in this hypothetical wallet's value over time, assuming no staking, no token accumulation, and no withdrawals.

$$Baseline\ Price = Price_{week\ =\ baseline\ week}$$

$$Initial\ Quantity\ = \ \frac{100}{Baseline\ Price}$$

---

[2] Imagine you invested $100 in a certain cryptocurrency on day one. The 'wallet value' tracks the value of that investment over time.

$$Wallet\ Value\ =\ Initial\ Quantity\ \times\ Price_{week}$$

There are 9 and 11 control variables in the 2024 and the 2020 models, respectively. The selected control variables are diverse indicators of holdings and transaction activity. They were chosen from a pool of all available predictors using a correlation matrix with a threshold of 0.7 – predictors correlated with another independent variable stronger than this were excluded.

| DV/IV | Variable | Meaning |
|-------|----------|---------|
| DV | Wallet Value | Tracks the value of a $100 wallet investment in the cryptocurrency made in the first week (2023-04-02) over time, representing the change in the 'wallet' value. |
| IV | Active Addresses Ratio | The ratio of active addresses to the total addresses with a balance, indicating network activity level as a percentage. |
| IV | Average Time Between Transactions | Measures the average interval in seconds between transactions, reflecting activity regularity and transactional demand on the network. |
| IV | Balance by Time Held: Traders | The number of traders (defined as wallet addresses holding a balance worth under 1 million dollars). |
| IV | Daily Active Addresses: Zero Balance Addresses (and the share of zero balance addresses out of total daily active addresses) | The number of addresses that have transferred out all of their tokens, indicating user exits from the network or short-term transactional behavior. |
| IV | Hodler Balance (and the share out of all categories by time held) | The aggregated balance of coins held by addresses classified as "Hodlers" (those holding coins for over one year), providing insight into long-term holding behavior. |
| IV | New Adoption Rate | The proportion of new addresses that made their first transaction out of all active addresses, highlighting the rate of new user adoption over time. |
| IV | Number of Addresses by Holdings worth $0-1 USD | The count of addresses holding balances within the $0–$1 USD range provides insights into tiny holders' participation in the network. |
| IV | Number of Addresses by Holdings worth $1-10 USD (and the share out of the total | The count of addresses holding balances within the $1–$10 USD range, providing insights into smaller holders' participation in the network. |
| | holdings) | |
| IV | Number of Addresses by Holdings worth $10-100 USD | The number of addresses holding between $10 and $100 in value, reflecting mid-level holders' participation. |
| IV | Number of All-Time Highers (and the log of it) | The count of addresses that bought within 20% of the cryptocurrency's all-time high price, indicating recent high-interest purchases. |
| IV | Number of All-Time Lowers (and the log of it) | The count of addresses that bought within 20% of the all-time low price, representing entries into the asset at its lowest historical levels. |
| IV | Number of Transactions (and the log of it) | The count of the number of valid transactions (excluding failed and reverted ones). |
| IV | Time Between Transactions Normalized by Total Addresses | The average time between transactions (in seconds), normalized by the number of total addresses for a given token (reflecting activity regularity and transactional demand on the network). |
| IV | Trader Balance (and the share out of all categories by time held) | The aggregated balance of coins held by addresses classified as "Traders" (those holding coins for less than one month), representing short-term holding and potentially speculative behavior. |
| IV | Total Addresses (and the normalized version)[3] | The number of addresses in the network. The normalized version creates a scale from 0 to 1 by dividing the observations by the maximum number of total addresses in the dataset for that cryptocurrency. |
| IV | Total Addresses: Zero Balance Addresses (and the share of zero balance addresses out of total addresses) | All the addresses that used to hold a crypto-asset but no longer do. |

**Table 2**. The list of variables used to build the synthetic control model to study the price impact of the 2024 and 2020 Bitcoin halvings. The first column indicates whether the variable is dependent (DV) or independent (IV).

**Donor Pool and Covariate Weights**

*2024 Halving*

---

[3] The normalization strategy was to divide the given observation from the maximum total addresses value for the given time period, thus scaling the variable to take values between 0 and 1.

The donor pool for constructing the synthetic Bitcoin is composed of a set of cryptocurrencies that share similar market dynamics but did not undergo a halving event during the study period. These cryptocurrencies serve as plausible comparison units, allowing for an estimation of Bitcoin's price trajectory absent the halving event by forming a "synthetic Bitcoin" based on a weighted average of these units. Table 3 presents the non-zero weighted cryptocurrencies contributing to the synthetic Bitcoin, derived from the optimization process, which minimizes the pre-treatment differences in the chosen predictors between Bitcoin and the synthetic control. Tron has the greatest contribution to the synthetic Bitcoin, with almost half of all the weights. The most influential predictor is the Active Addresses Ratio. Therefore, the equation for the synthetic control unit is:

Synthetic Bitcoin = 0.351 * Active Addresses Ratio + 0.004 * Share of the Number of Addresses by Holdings worth $1-10 USD + 0.02 * Log of the Number of All-Time Highers + 0.064 * Log of the Number of All-Time Lowers + 0.001 * Share of Zero Balance Addresses (from Daily Active Addresses) + 0.113 * Share of Hodler Balance Addresses + 0.239 * The Share of Trader Balance Addresses +  0.099 * New Adoption Rate + 0.11 * Share of Zero Balance Addresses (from Total Addresses)

| Cryptocurrency | Abbreviation | Weight | Predictor | Weight |
|---|---|---|---|---|
| **Tron** | TRX | 0.423 | **Active Addresses Ratio** | 0.351 |
| **Cardano** | ADA | 0.251 | **Traders Balance Share** | 0.239 |
| **MEXC Exchange Token** | MX | 0.154 | **Hodler Balance Share** | 0.113 |
| **Fetch.AI** | FET | 0.109 | **Total Zero Balance Share (from Daily Active Addresses)** | 0.11 |
| **KuCoin** | KCS | 0.063 | **New Adoption Rate** | 0.099 |
| | | | **Number of All-Time Lowers** | 0.064 |

| | | | (Log) | |
|---|---|---|---|---|
| | | | **Number of All-Time Highers (Log)** | 0.02 |
| | | | **Share of the Number of Addresses by Holdings worth $1-10 USD** | 0.004 |
| | | | **The Share of Zero Balance Addresses (from Total Addresses)** | 0.001 |

**Table 3**. 2024 Model Donor Pool Cryptocurrencies and Their Corresponding Synthetic Control Weights.

Table 4 shows the pre-intervention characteristics of Bitcoin, its synthetic counterpart, and an average of 23 cryptocurrencies in the donor pool. The SCM model obtains a closer match for six out of nine predictors compared to the sample mean. For New Adoption Rate and Zero Balance Addresses, the treated value is almost double the synthetic one. For other covariates, the differences between the treated and synthetic covariate values are smaller. In the following section, I estimate the causal effect of 2024 Bitcoin halving on its price as the difference in the wallet value between Bitcoin and its synthetic counterpart after April 20, 2024.

| Predictor | Treated | Synthetic | Sample mean | Improvement |
|---|---|---|---|---|
| **Active Addresses Ratio** | 0.019 | 0.014 | 0.006 | 0.008 |
| **Traders Balance Share** | 0.078 | 0.081 | 0.064 | 0.011 |
| **Hodler Balance Share** | 0.688 | 0.591 | 0.609 | -0.018 |
| **Total Zero Balance Share (from Daily Active Addresses)** | 0.486 | 0.239 | 0.271 | -0.032 |
| **New Adoption Rate** | 0.484 | 0.251 | 0.304 | -0.053 |
| **Number of All-Time Lowers (Log)** | 10.321 | 7.911 | 7.069 | 0.842 |
| **Number of All-Time Highers (Log)** | 13.631 | 9.512 | 7.747 | 1.765 |
| **Share of the Number of Addresses by Holdings worth $1-10 USD** | 23.424 | 16.873 | 14.369 | 2.504 |
| **The Share of Zero Balance Addresses (from Total Addresses)** | 0.96 | 0.69 | 0.616 | 0.074 |

**Table 4**. Comparison of pre-treatment predictor values for the treated, the synthetic control unit, and all the units in the 2024 model sample. The column on the right shows the improvement in the absolute difference between the treated and synthetic pre-treatment predictor values compared to the absolute distance between the treated and sample mean values (or improvement in the distance from the treated pre-treatment predictor value).

*2020 Halving*

The obtained donor and predictor weights for the 2020 halving SCM model are shown in Table 5, following the same process. KuCoin has the greatest contribution to the synthetic Bitcoin, with over 50% of all the weights. Ten out of eleven predictors received a weight, and the normalized trader balance received the highest one. The equation for the weighted synthetic control unit is therefore:

Synthetic Bitcoin = 0.528 * Trader Balance (Normalized) + 0.223 * Total Addresses (Normalized) + 0.101 * The Share of Zero Balance Addresses (from Total Addresses) + 0.078 * Number of Addresses by Holdings worth $10k-100k USD (share) + 0.042 * New Adoption Rate + 0.013 * Active Addresses Ratio + 0.006 * Number of Addresses by Holdings worth $1-10 USD (share) + 0.004 * Log of the Number of Transactions + 0.003 * Time Between Transactions (normalized by Addresses) + 0.001 * Number of Addresses by Holdings worth $0-1 USD (share) + 0 * Number of Addresses by Holdings worth $10-100 USD (share)

| Cryptocurrency | Abbreviation | Weight | Predictor | Weight |
|---|---|---|---|---|
| **KuCoin** | KCS | 0.564 | **Trader Balance (Normalized)** | 0.528 |
| **Cardano** | ADA | 0.211 | **Total Addresses (Normalized)** | 0.223 |
| **Fantom** | FTM | 0.114 | **The Share of Zero Balance Addresses (from Total Addresses)** | 0.101 |
| **Chainlink** | LINK | 0.109 | **Number of Addresses by Holdings worth $10k-100k USD (share)** | 0.078 |

| Dogecoin | DOGE | 0.001 | New Adoption Rate | 0.042 |
|---|---|---|---|---|
| | | | Active Addresses Ratio | 0.013 |
| | | | Number of Addresses by Holdings worth $1-10 USD (share) | 0.006 |
| | | | Log of the Number of Transactions | 0.004 |
| | | | Time Between Transactions (normalized by Addresses) | 0.003 |
| | | | Number of Addresses by Holdings worth $0-1 USD (share) | 0.001 |
| | | | Number of Addresses by Holdings worth $10-100 USD (share) | 0 |

**Table 5**. 2020 Model Donor Pool Cryptocurrencies and Their Corresponding Synthetic Control Weights.

Table 6 compares the pre-intervention characteristics of Bitcoin, its synthetic counterpart, and an average of 19 cryptocurrencies in the donor pool. Ethereum was removed from the donor pool due to significant idiosyncratic shocks to its price between February 2023 and July 2024 – the after-effects of the Merge, the Shanghai and Cancun-Deneb upgrades, and the spot Ethereum ETF approval by the SEC. These likely interacted with the BTC halving and would probably not have affected the ETH price to the same degree in the absence of the halving. The SCM model obtains a closer match for five out of eleven predictors compared to the sample mean. Except for the Total Addresses (Normalized), the differences between the treated and synthetic characteristics are not large. In the following section, I estimate the causal effect of 2020 Bitcoin halving on its price as the difference in the wallet value between Bitcoin and its synthetic counterpart after May 11, 2020.

| Predictor | Treated | Synthetic | Sample mean | Improvement |
|---|---|---|---|---|
| Trader Balance (Normalized) | 0 | 0.019 | 0.036 | 0.017 |
| Total Addresses (Normalized) | 0.588 | 0.001 | 0.003 | -0.002 |
| The Share of Zero Balance Addresses (from Total Addresses) | 0.954 | 0.752 | 0.613 | 0.139 |
| Number of Addresses by Holdings worth $10k-100k USD (share) | 1.824 | 2.058 | 3.336 | 1.278 |
| New Adoption Rate | 0.508 | 0.358 | 0.278 | 0.080 |
| Active Addresses Ratio | 0.029 | 0.011 | 0.037 | -0.010 |
| Number of Addresses by Holdings worth $1-10 USD (share) | 23.221 | 4.640 | 14.717 | -10.077 |
| Time Between Transactions (normalized by Addresses) | 0 | 0.031 | 0.117 | 0.086 |
| Log of the Number of Transactions | 12.672 | 5.460 | 6.33 | -0.870 |
| Number of Addresses by Holdings worth $0-1 USD (share) | 27.552 | 63.060 | 34.792 | -28.268 |
| Number of Addresses by Holdings worth $10-100 USD (share) | 24.679 | 8.637 | 19.043 | -10.406 |

**Table 6**. Comparison of pre-treatment predictor values for the treated, the synthetic control unit, and all the units in the 2020 model sample. The column on the right shows the improvement in the absolute difference between the treated and synthetic pre-treatment predictor values compared to the absolute distance between the treated and sample mean values (or improvement in the distance from the treated pre-treatment predictor value).

**Analysis**

*2024 Halving*

Figure 3 shows the wallet value for Bitcoin and its synthetic counterpart from the week of April 2nd, 2023 (denoted as week 1 in the code) until the week of July 21st, 2024 (week 68). The outcome variable trend for the synthetic Bitcoin closely follows that of the actual Bitcoin for the whole pre-treatment period. This trend alignment, alongside the relatively well-distributed weights among the wallet value predictors (Table 3, with a standard deviation of 0.12), offers evidence that the constructed synthetic Bitcoin could be a reliable estimation of the wallet value we would have observed if the 2024 Bitcoin halving did not occur (for instance, if the community collectively decided to skip this halving through a majority vote). Thus, the treatment effect of Bitcoin halving on the wallet value (as a proxy of price) is the difference between the post-intervention wallet value trends for Bitcoin and the synthetic Bitcoin. Figure 3 suggests that following the halving, the two trends diverge, displaying a predominantly positive treatment effect.

Figure 4 shows the weekly estimates of the impact of Bitcoin halving on its price. It fluctuates between April 20, 2024, and July 21st, 2024, with an average of $24.55. This could be interpreted as follows: assuming an investor purchased $100 worth of Bitcoin in the week of April 2nd, 2023, the value of this investment (a proxy of price) would be $24.55 greater on average within the next 3 months after halving as a result of the 2024 Bitcoin halving, controlling for other factors and relying on the SCM assumptions. Since the Wallet Value (the Y-variable) started tracking the change in cryptocurrency price with a hypothetical $100 initial investment, this gap corresponds to roughly a 24.55% higher return compared to a counterfactual scenario in which the halving did not occur. However, one significant concern of this study is that the Bitcoin halving had spillover effects, influencing the prices of other cryptocurrencies in the donor pool since all of the altcoins in the pool are known to be correlated with Bitcoin. Closely evaluating the possibility of such spillovers is crucial. A manual Google Search for major shocks (e.g., upgrades, major hacks, chain mergers) during this period among the donor pool units was conducted, but nothing of major concern was discovered.

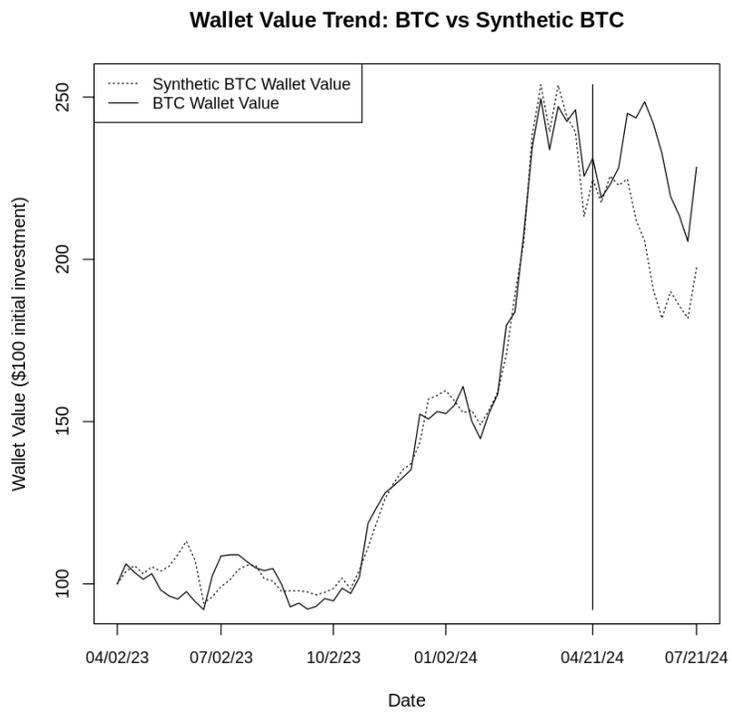

**Figure 3.** Wallet Value from April 2nd, 2023, until July 21st, 2024: Bitcoin versus synthetic Bitcoin. The vertical dotted line shows the week of the 2024 Bitcoin halving.

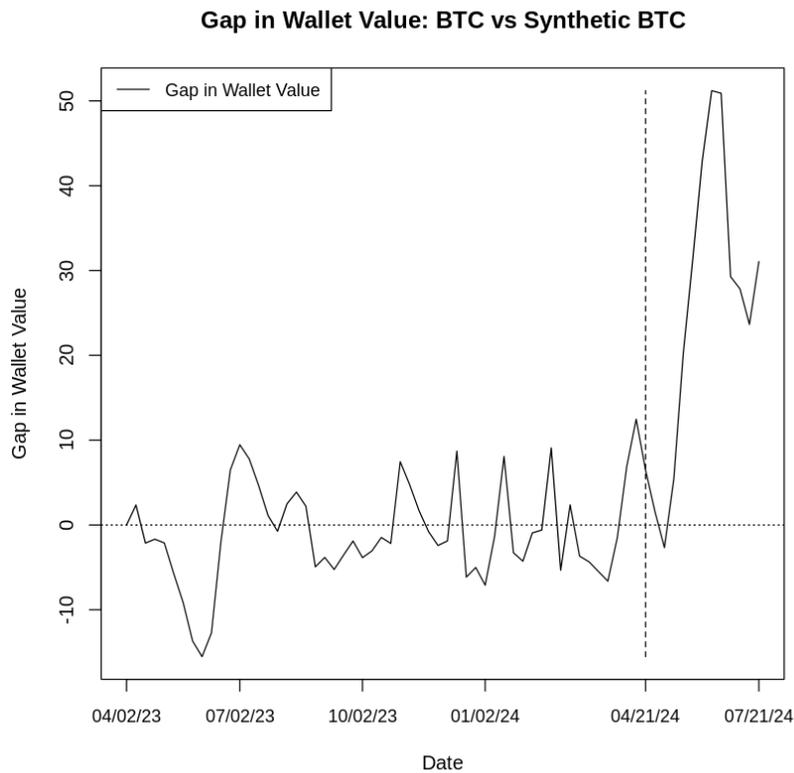

**Figure 4.** Wallet Value Gap between Bitcoin (continuous line) and synthetic Bitcoin (dotted horizontal line) from April 2nd, 2023 until July 21st, 2024. The vertical dotted line shows the week of the 2024 Bitcoin halving.

*2024 Halving: Placebo Tests*

To verify the reliability of the causal estimates mentioned in the previous section, I conduct placebo tests. The goal of the tests is to understand whether the outcome variable trend after halving can, in fact, be attributed to the halving as opposed to chance. Placebo tests are particularly crucial for verifying results obtained from an SCM model because the model is blind to the treatment itself. Passing the model the treatment time as the input, we hope that it captures the right treatment, but it is not guaranteed.

First, I conduct a placebo test in time: I shift the treatment time 6 months (24 weeks) back to see if the SCM still finds a consistent post-treatment divergence (or, in other words, we want to see if the results did not change). Such long backtracking is chosen intentionally to attempt to place the 'placebo' treatment time prior to the moment when the expectations of the 2024 Bitcoin halving begin materializing (Figure 5). Everything else in the model stays constant.

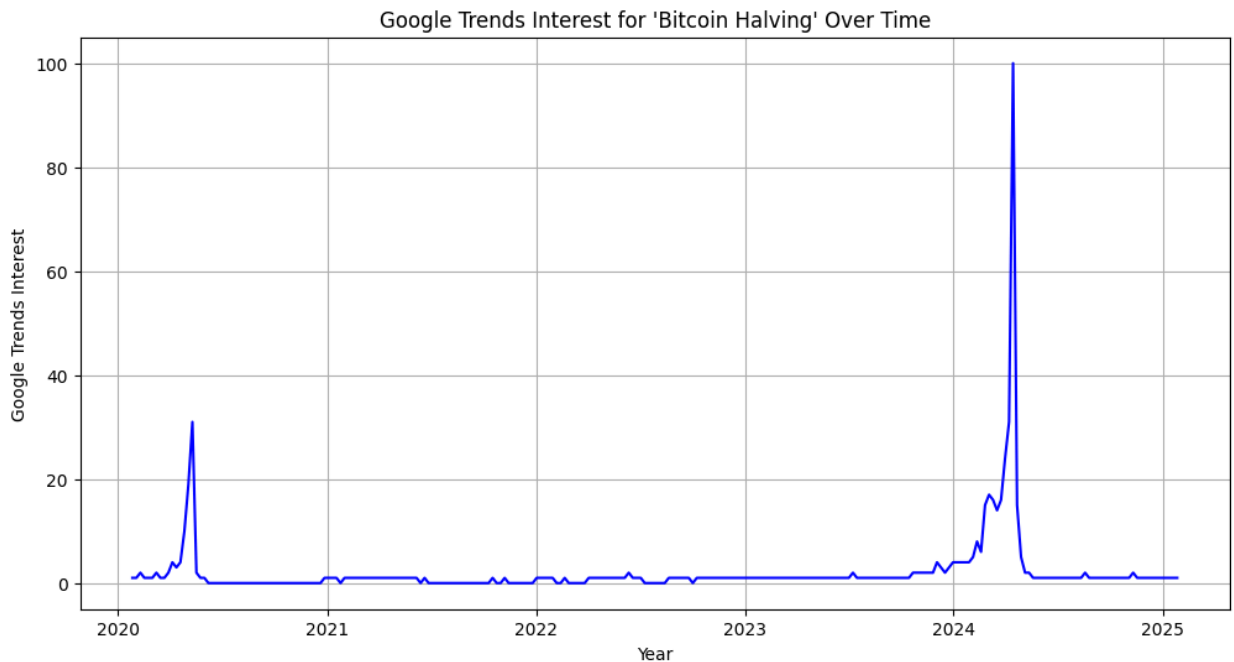

**Figure 5.** This graph shows the Google Trends relative search interest for 'Bitcoin Halving' from 2020 to 2025. Numbers represent search interest relative to the highest point on the chart for the time period worldwide. A value of 100 is the peak popularity for the term. A score of 0 means there was not enough data for this term. The smaller peak in 2020 is the 2020 Bitcoin halving. Since the 2020 Bitcoin halving, the relative interest seems to have grown consistently at the beginning of November of 2023, which is about 24 weeks before the 2024 halving date.

Notably, the wallet value trajectory began diverging in the week of March 3rd, 2024 (~50 days prior to the Bitcoin halving). Normally, this would provide enough evidence to conclude that the in-time placebo test was not passed successfully. However, in the context of such a highly expected event, as discussed in the Literature Review, this post-treatment wallet value divergence could be attributed to 'animal spirits' – investors' expectations of halving driving the price up. In other words, this can be called the "50-days prior effect."

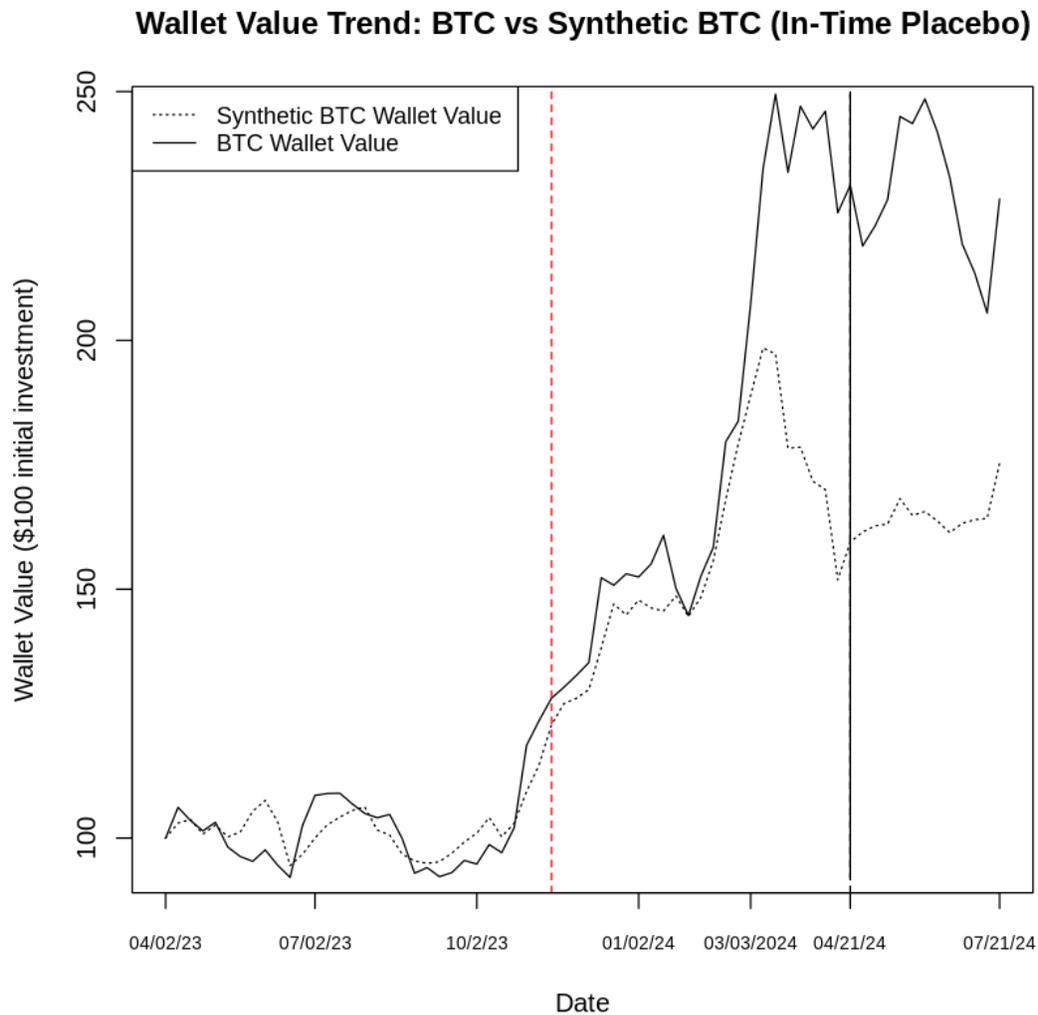

**Figure 6.** Wallet value from April 2nd, 2023 (denoted as week 1) until July 21st, 2024 (week 68). For this placebo test, the treatment is shifted 24 weeks back, from the week of April 20, 2024, to the week of November 4, 2023. The black vertical line shows the week of the 2024 Bitcoin halving, and the dotted red line shows the new placebo treatment time.

I also conduct an in-space placebo study, swapping out the treated unit in the SCM model

for Ankr, a donor unit that, in theory, should not be impacted by the Bitcoin halving. Figure 7

does not display an ideal pre-treatment fit starting from the end of 2023, but the post-treatment

Wallet Value trend is almost identical. Therefore, this placebo test is passed, providing additional evidence that the SCM model is likely capturing the intended treatment – the 2024 Bitcoin halving.

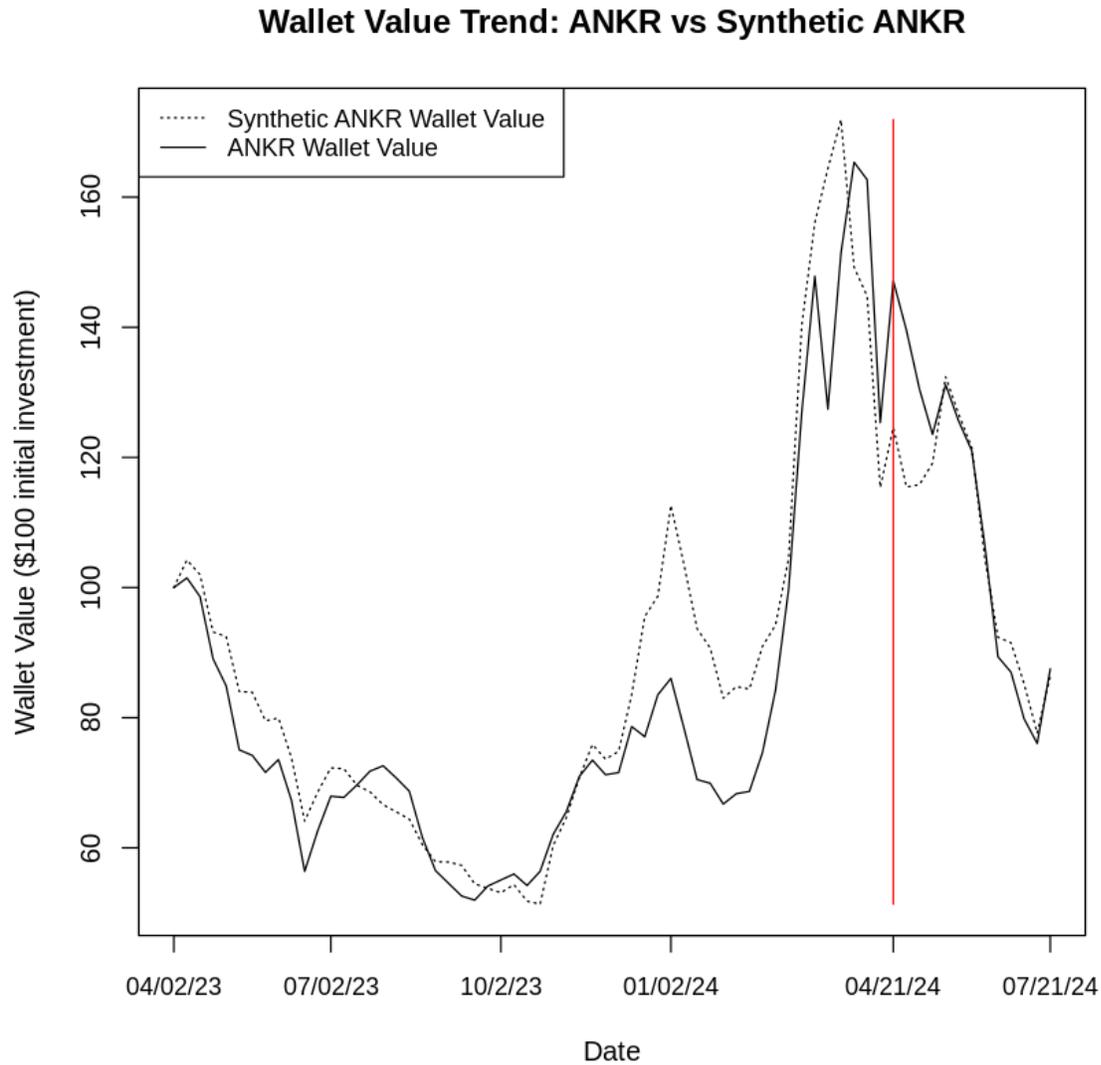

**Figure 7.** Wallet value from April 2nd, 2023 (denoted as week 1) until July 21st, 2024 (week 68). The treatment unit is Ankr. The red vertical line shows the time of the 2024 Bitcoin halving. Notably, the post-treatment divergence is insignificant.

Finally, I repeat this process and iteratively reassign the treatment to all comparison units in my sample. This way, I obtain SCM estimates for all other cryptocurrencies that did not undergo the halving. Then, I compare the effect of the Bitcoin halving on the wallet value (a proxy of price) to the distribution of these placebo effects. The Bitcoin halving effect can be considered substantial (unusually large) if that effect is significantly different from the distribution of gaps obtained from the placebo tests.

Figure 8 shows the outcome of this placebo test, excluding cryptocurrencies with a pre-intervention MSPE larger than 10 times that of Bitcoin. The gray lines show the difference (gap) in wallet value between each donor cryptocurrency and its synthetic counterpart. The black line shows the original gap for Bitcoin.

Since this gap after the Bitcoin halving is greater for Bitcoin than the gaps for the cryptocurrencies in the donor pool, this placebo test is passed successfully, yielding additional evidence that the treatment effect is not due to chance. Results of the same placebo test with a larger cutoff are presented in Appendix 6.

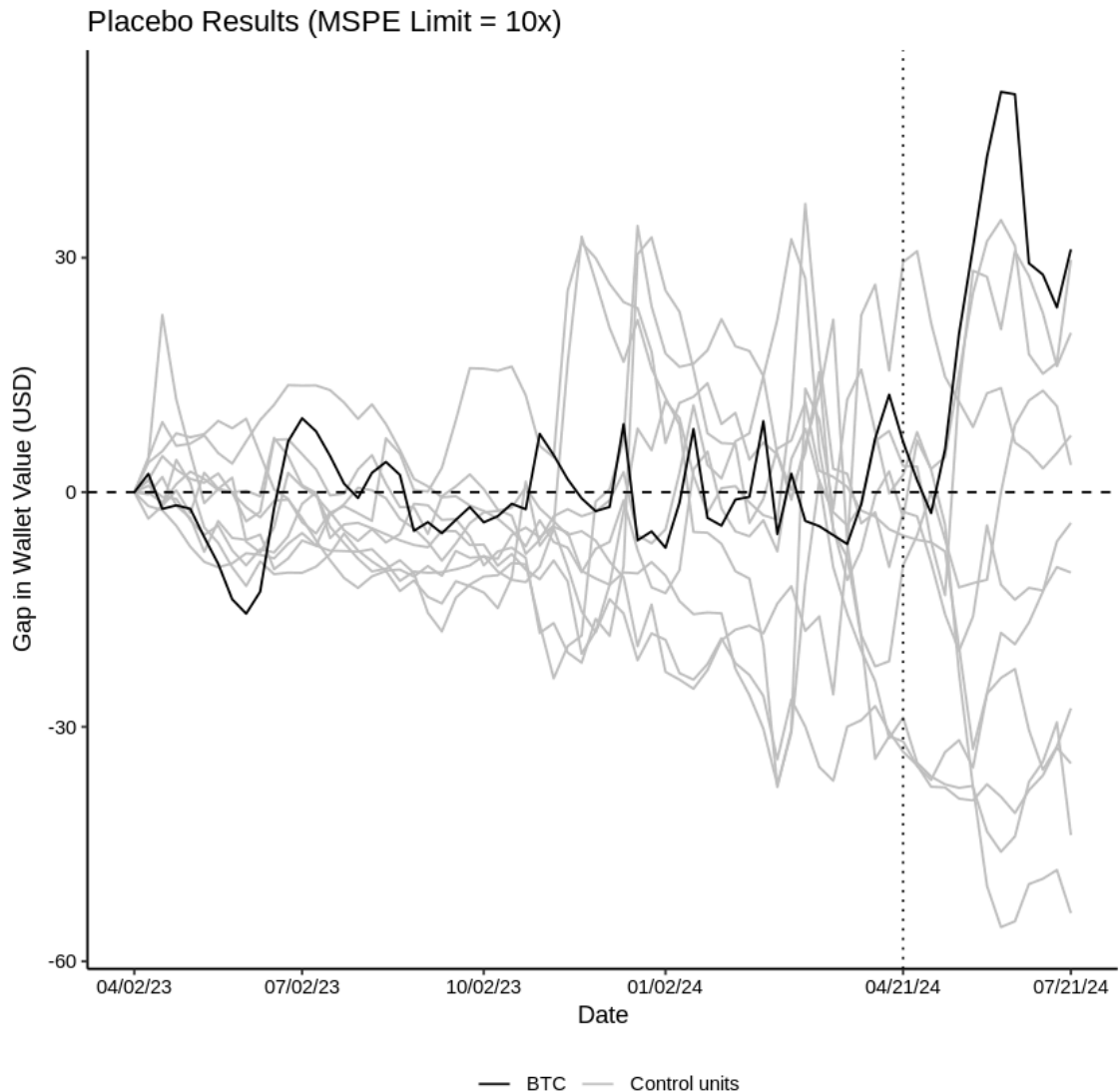

**Figure 8**. This plot shows the difference between the observed unit and synthetic controls for the treated and control units, in black and grey, respectively. 13 cryptocurrencies with pre-Halving Mean Squared Prediction Error (MSPE) 10 times higher than Bitcoin's are discarded (ALGO, FET, FTM, GNO, IOTX, LEO, LINK, LPT, MATIC, MKR, MX, QNT, and TRX). Wallet value gaps for Bitcoin and 10 remaining placebo gaps for control cryptocurrencies are shown. The vertical dotted line shows the time of the 2024 Bitcoin halving.

I also evaluate the gap by examining the distribution of post- and pre-halving mean squared prediction error (MSPE) ratios. MSPE is an indicator of discrepancy in the wallet value between each cryptocurrency and its synthetic counterpart. Figure 9 shows these ratios for Bitcoin and other cryptocurrencies in the donor pool. Notably, Bitcoin has the highest post/pre-MSPE ratio (close to 30). Livepeer's (LPT) ratio is the second highest – the post-halving gap is about 22 times greater than the pre-halving gap. Thus, the probability of randomly selecting a cryptocurrency with a ratio as high as Bitcoin's is 1/24 or 0.0416, which is under the 0.05 threshold and can be considered statistically significant.

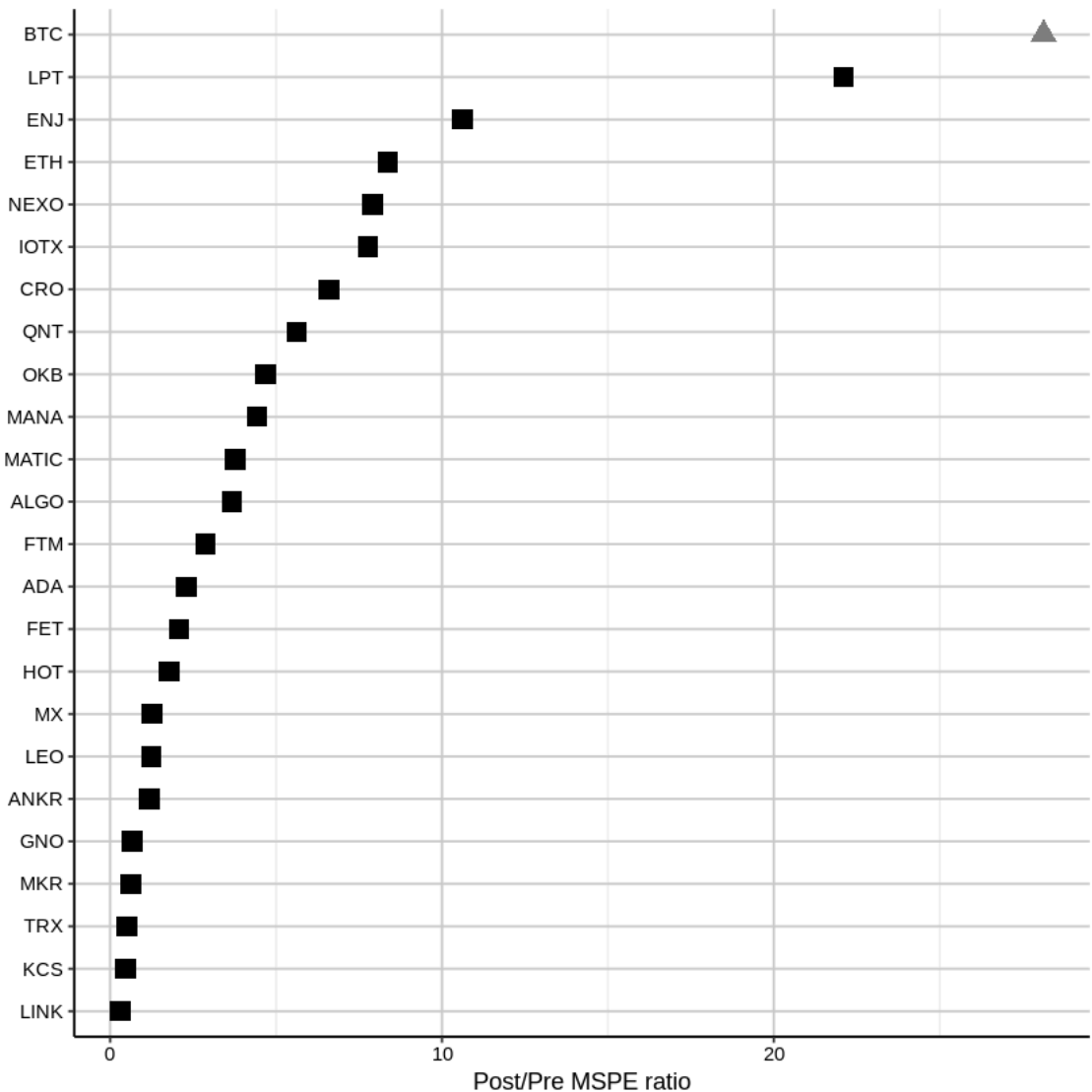

**Figure 9.** The ratio of Post-halving MSPE to pre-crisis MSPE: Bitcoin and Control Cryptocurrencies.

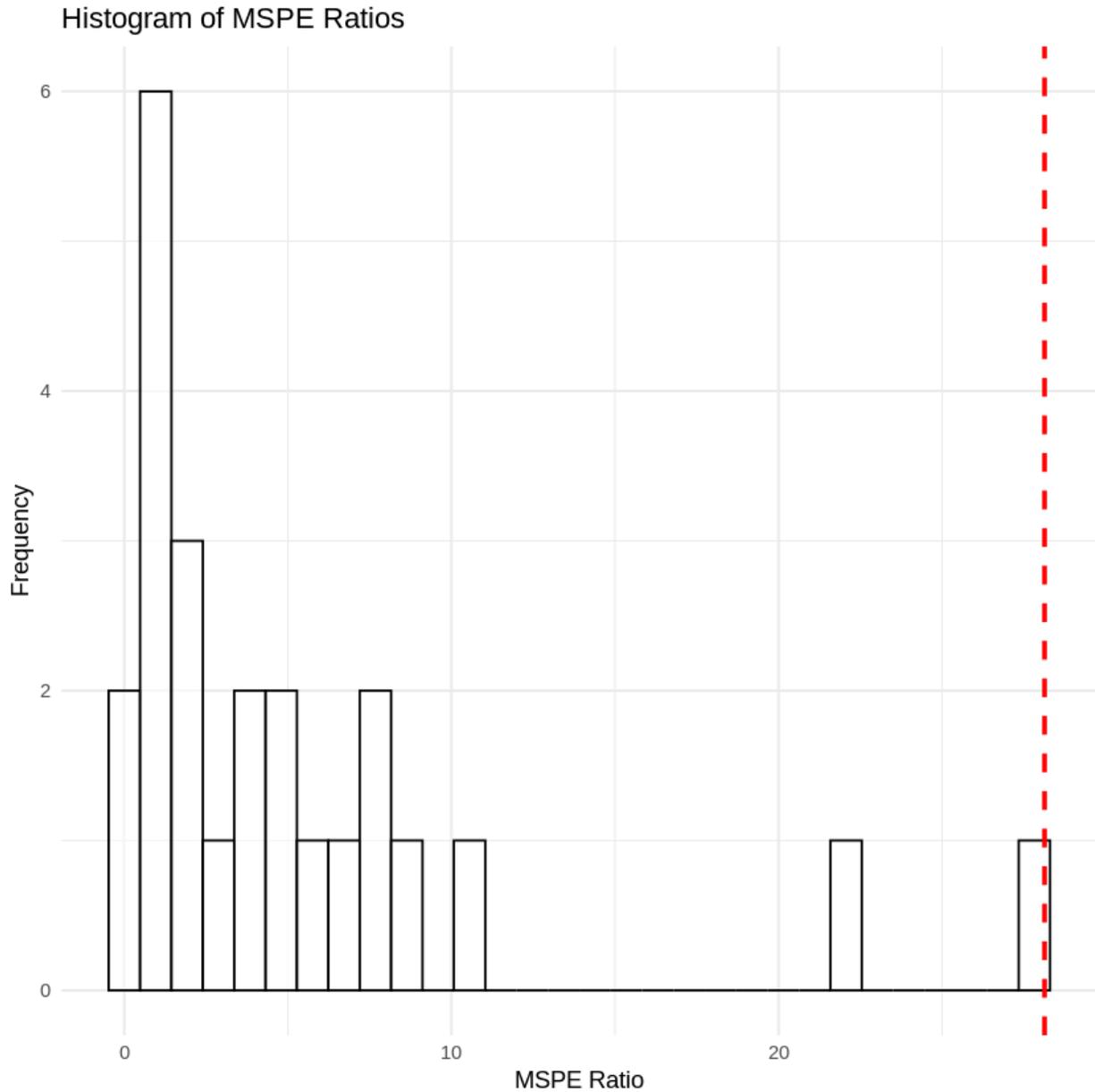

**Figure 9.1.** The histogram shows the ratio of the Post-halving MSPE to pre-crisis MSPE: Bitcoin and Control Cryptocurrencies. The column on the very right corresponds to the ratio of Bitcoin.

*2020 Halving*

Figure 10 shows the wallet value for Bitcoin and its synthetic counterpart from May 5th,

2019 (denoted as week 1 in the code) until August 9, 2020 (week 66). Similarly to the previous model, the seemingly good enough pre-treatment outcome variable trend alignment, alongside the well-distributed weights among the wallet value predictors (Table 5, with a higher standard deviation of 0.16), offers evidence that the constructed synthetic Bitcoin could be a reliable estimation of the wallet value we would observe if the 2020 Bitcoin halving did not occur.

Figures 10 and 11 indicate a positive effect during the first three weeks, which then reverses and becomes increasingly negative, reaching -$95.5. This contrasts with the findings from the 2024 halving study. The average effect between the weeks of May 11, 2020, and August 9, 2020, is -$29.1 in Wallet Value. This could be interpreted as follows: assuming an investor purchased $100 worth of Bitcoin during the week of May 5th, 2019, the value of this investment (a proxy of price) would be $29.1 lower on average within the next 3 months after halving as a result of the 2020 Bitcoin halving, controlling for other factors and relying on the SCM assumptions. Since we started with a $100 investment, this gap corresponds to roughly a 29.1% lower return compared to a counterfactual scenario in which the halving didn't occur.

However, one significant concern of this study is that the SCM model is picking up a treatment other than the well-anticipated 2020 Bitcoin halving. There is also reason to believe that the assumption of no interference is violated since the donor units' growth was likely positively influenced by the bitcoin halving. If that is the case, then this estimated treatment effect for Bitcoin is biased downward. The reason is that synthetic control assumes that the donor pool represents what would have happened to Bitcoin if no halving had occurred. But if the donors themselves receive spillover gains (e.g., because broader crypto sentiment lifted altcoins after the halving), their post‑treatment outcome is artificially higher than a proper "untreated" baseline. As a result, comparing Bitcoin's actual outcome to that elevated donor outcome, a spuriously negative treatment effect could be observed. Finally, this estimate could be biased because the research design exploits time variation in the Wallet Value to estimate the effect of a very well-anticipated Bitcoin halving. Abadie (2021) warns that "if forward-looking economic agents react in advance of the policy intervention under investigation," the SCM estimators may be biased.

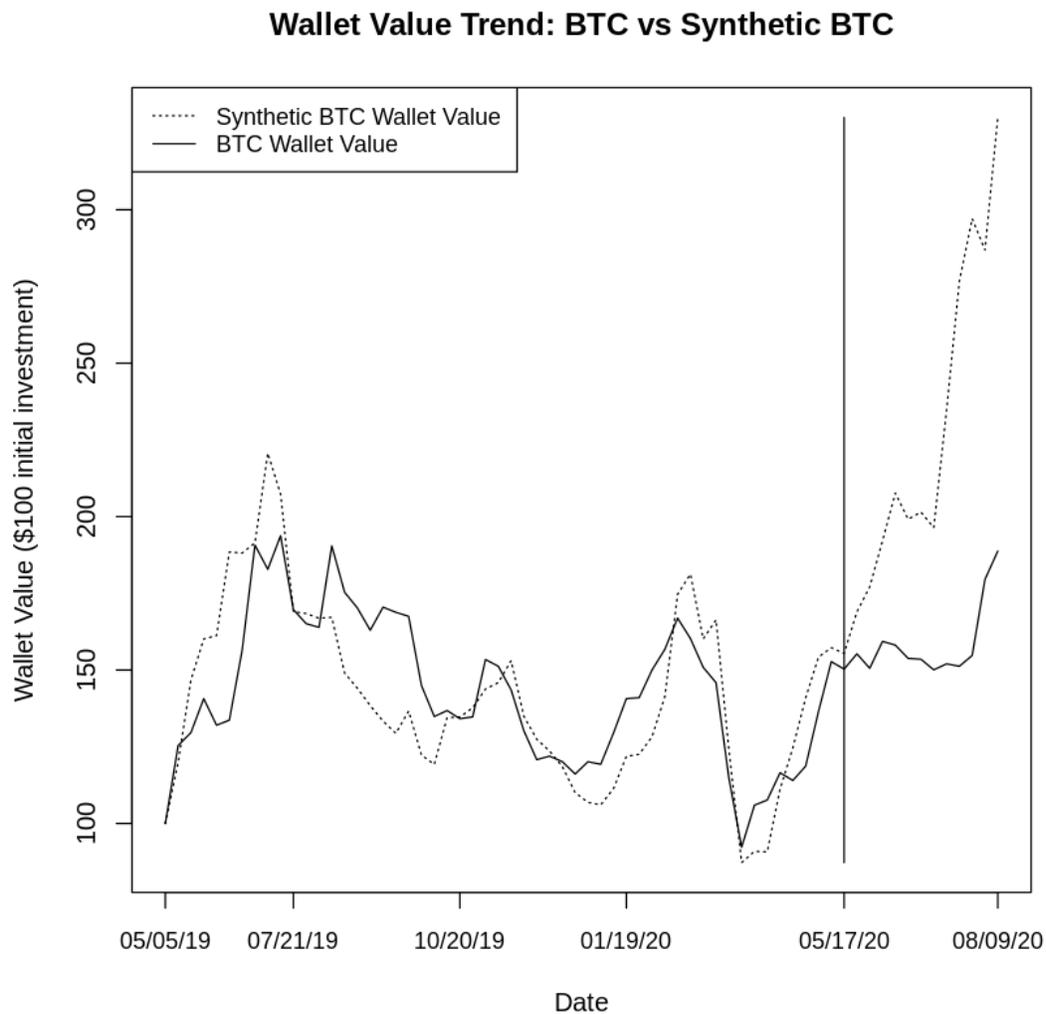

**Figure 10.** Wallet Value from the week of May 5th, 2019 (denoted as week 1) until the week of August 9, 2020 (week 66): Bitcoin versus synthetic Bitcoin. The vertical dotted line shows the week of the 2020 Bitcoin halving.

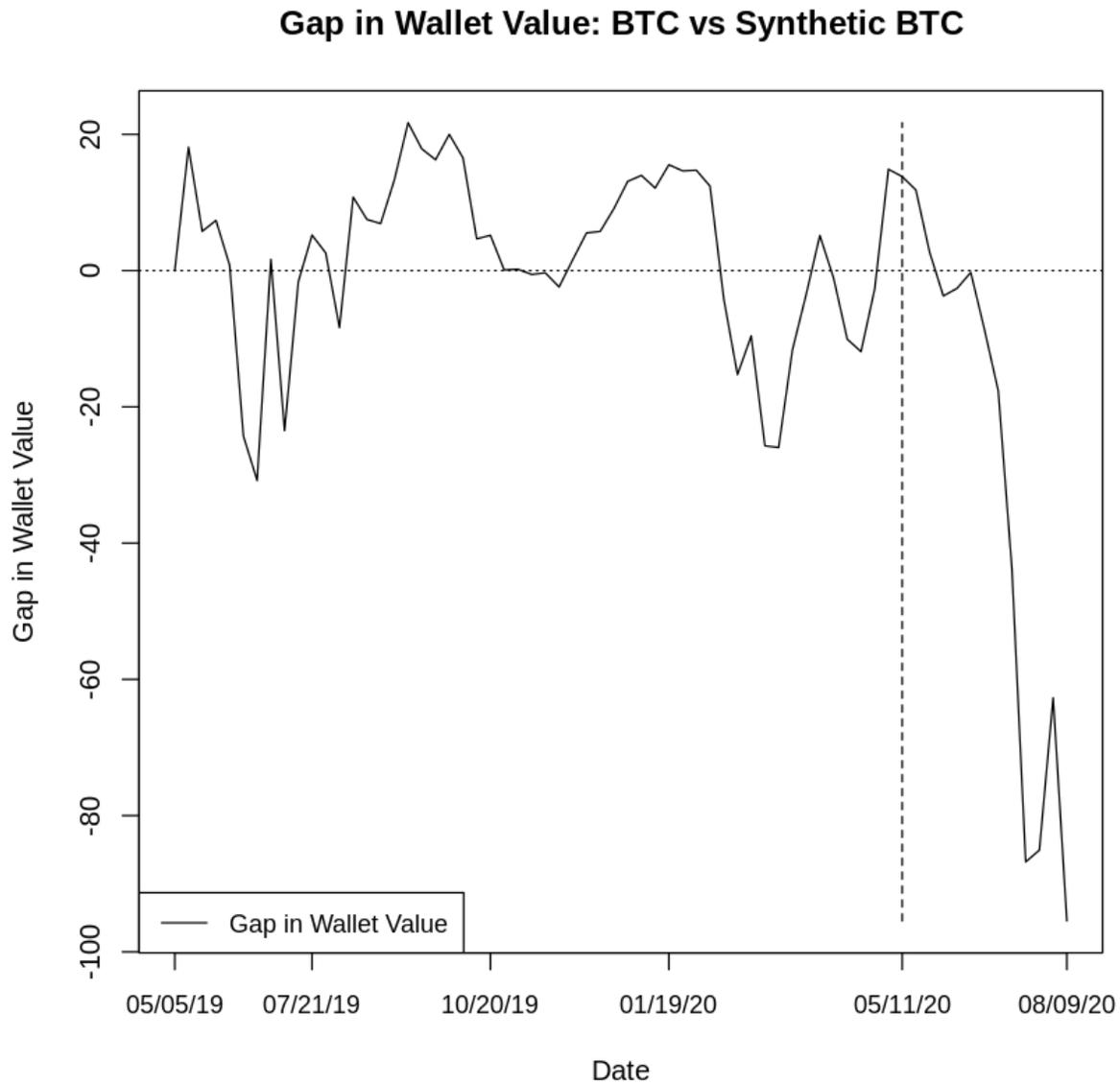

**Figure 11.** Wallet Value Gap between Bitcoin (continuous line) and synthetic Bitcoin (dotted horizontal line) from the week of May 5th, 2019 (denoted as week 1) until the week of August 9, 2020 (week 66). The vertical dotted line shows the time of the 2020 Bitcoin halving.

*2020 Halving: Placebo Tests*

Given the signs of anticipation of the Bitcoin halving discussed in the previous section, I

conduct a placebo in time – just like for the 2024 SCM model, I shift the treatment time 5

months (21 weeks) back to see if it still finds a consistent post-treatment divergence. Such long

backtracking is once again chosen intentionally to attempt to place the 'placebo' treatment time

prior to the moment when the expectations of the 2024 Bitcoin halving begin materializing

(Figure 12). Everything else in the model stays constant.

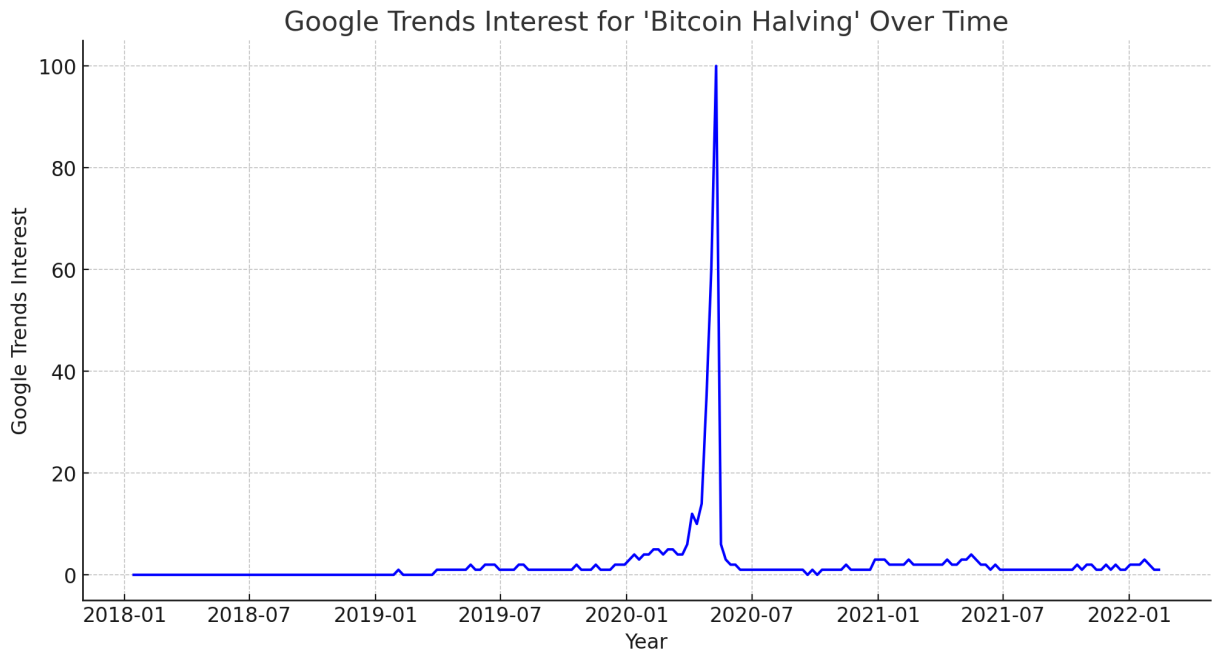

**Figure 12**. This graph shows the Google Trends relative search interest for 'Bitcoin Halving'
from 2018 to 2022. Numbers represent search interest relative to the highest point on the chart
for the time period worldwide. A value of 100 is the peak popularity for the term. A score of 0
means there was not enough data for this term. Since the 2020 Bitcoin halving, the relative
interest seems to have grown steadily in mid-December of 2019, which is about 21 weeks before
the 2024 halving date.

Figure 13 shows that the wallet value trajectory began diverging in the week of

December 15th, 2020, 21 weeks prior to the Bitcoin halving. Therefore, this in-time placebo test

is not passed successfully. The SCM model might be capturing the onset of the COVID-19

pandemic and the macroeconomic patterns associated with it or something else.

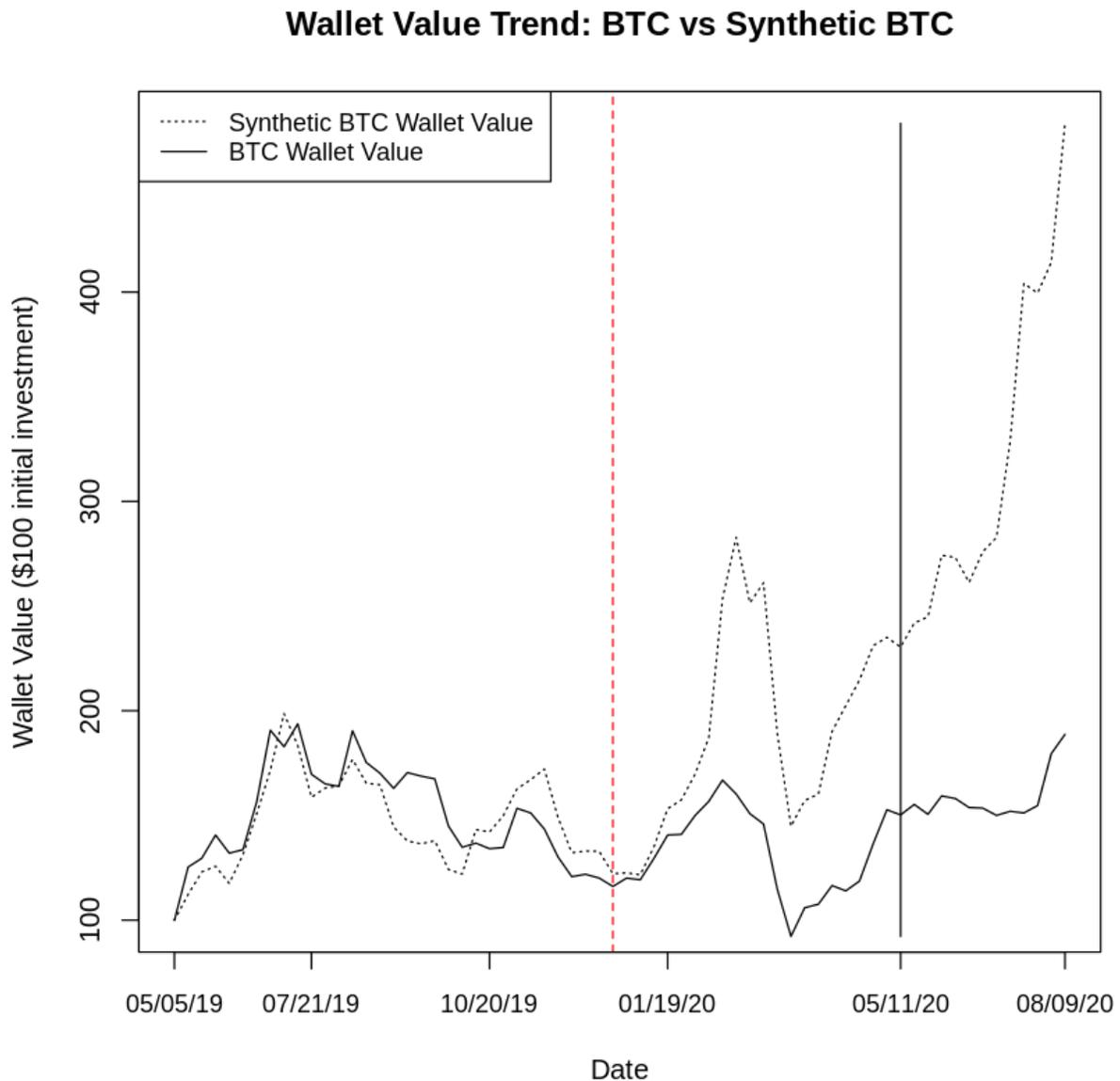

**Figure 13.** Wallet Value from May 5th, 2019 (denoted as week 1) until August 9, 2020 (week 66). For this placebo test, the treatment is shifted 21 weeks back to the week of December 15th, 2020, from week 54 to week 33. The black vertical line shows the week of the 2020 Bitcoin halving, and the dotted red line shows the new placebo treatment time.

I also conduct an in-space placebo test, selecting Cardano (ADA) as the treated unit because it had the second-largest donor weight in the 2020 SCM model. Figure 14 shows a negative treatment effect that flips positive after about three weeks. There is an increasing post-treatment Y divergence, indicating that this placebo test has also fails. In this model, Bitcoin's weight is 0.005. The model could be capturing coin-specific factors or broader market dynamics like the 2020 DeFi boom. Such a stark difference from the original treatment effect and its direction suggests that the model is not isolating the effect of halving cleanly.

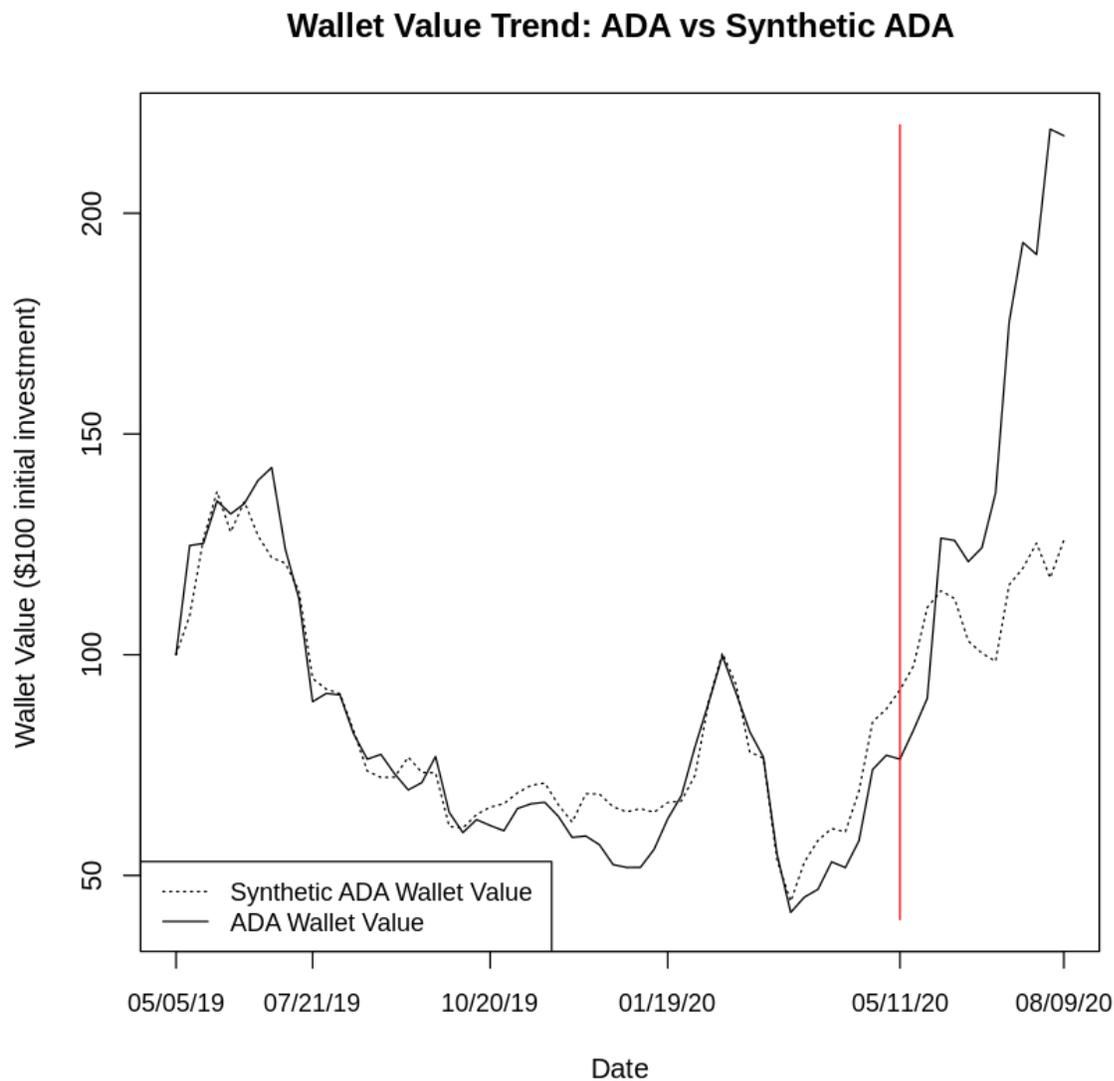

**Figure 14.** Wallet Value from May 5th, 2019 (denoted as week 1) until August 9, 2020 (week 66). The treatment unit is Cardano (ADA). The red vertical line shows the week of the 2020 Bitcoin halving.

Finally, I obtain SCM estimates for all other 19 cryptocurrencies that did not experience the 2020 halving directly. Figure 15 shows the results – the gray lines show the difference (gap) in wallet value between each donor cryptocurrency and its synthetic counterpart. The black line shows the original gap for Bitcoin. Since this gap after the Bitcoin halving is smaller than the gaps for the cryptocurrencies in the donor pool, this placebo test is not passed, and the obtained treatment effect could be due to chance.

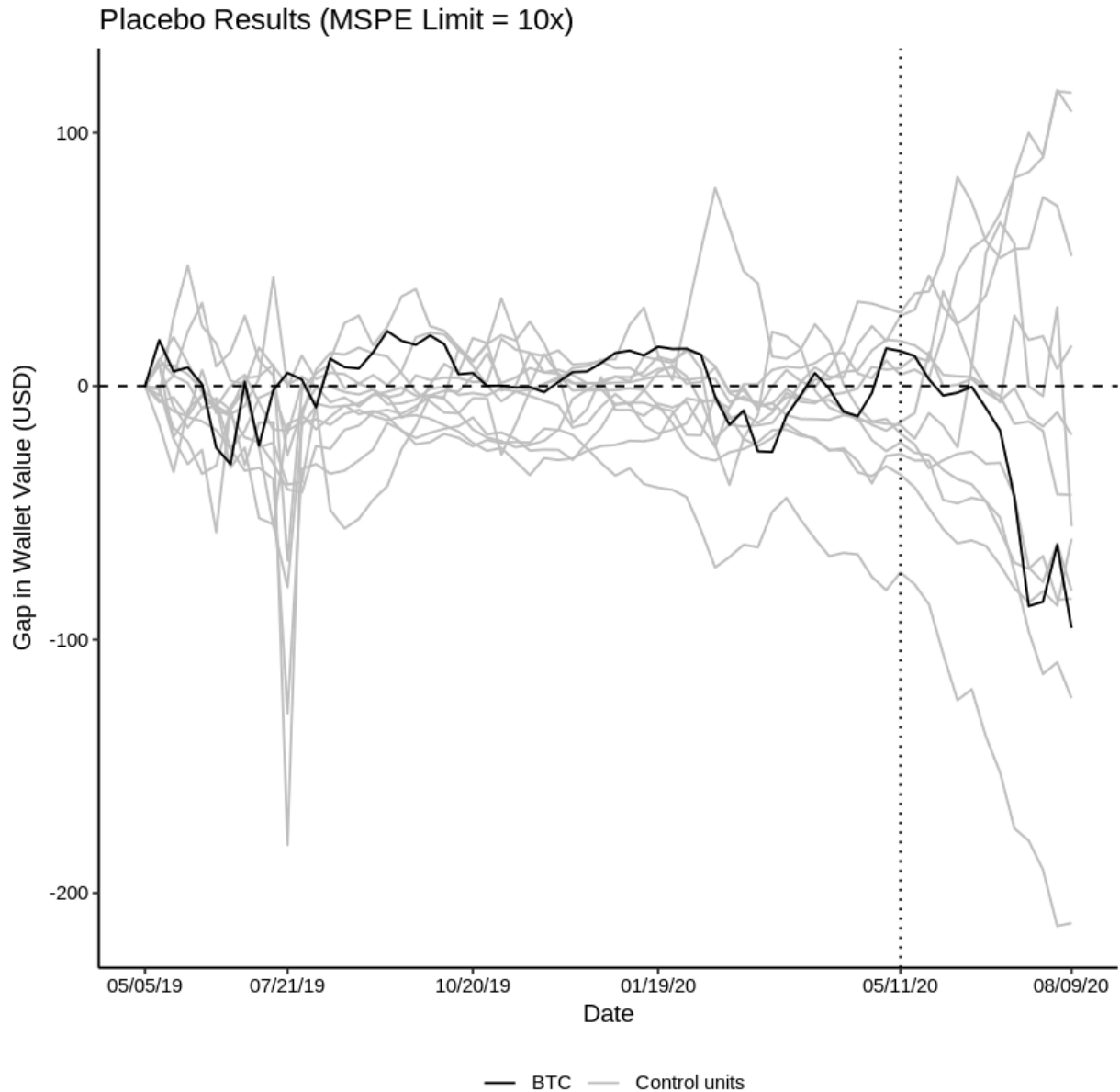

**Figure 15.** This plot shows the difference between the observed unit and synthetic controls for the treated and control units, in black and grey, respectively. 7 cryptocurrencies with pre-Halving Mean Squared Prediction Error (MSPE) 10 times higher than Bitcoin's are discarded (ANKR, FTM, LINK, MATIC, MKR, OKB, and QNT). Wallet value gaps for Bitcoin and 12 remaining placebo gaps for control cryptocurrencies are shown. The vertical dotted line shows the time of the 2020 Bitcoin halving.

I confirm the results shown in Figure 15 by evaluating the distribution of post- and pre-halving mean squared prediction error (MSPE) ratios. Figure 16 shows these ratios for Bitcoin and 19 other cryptocurrencies in the donor pool. Unsurprisingly, three control units have a higher post/pre-MSPE ratio than Bitcoin. Thus, the probability of randomly selecting a cryptocurrency with a ratio as high as Bitcoin's is 4/20 or 0.2. Therefore, the obtained treatment effect is not statistically significant.

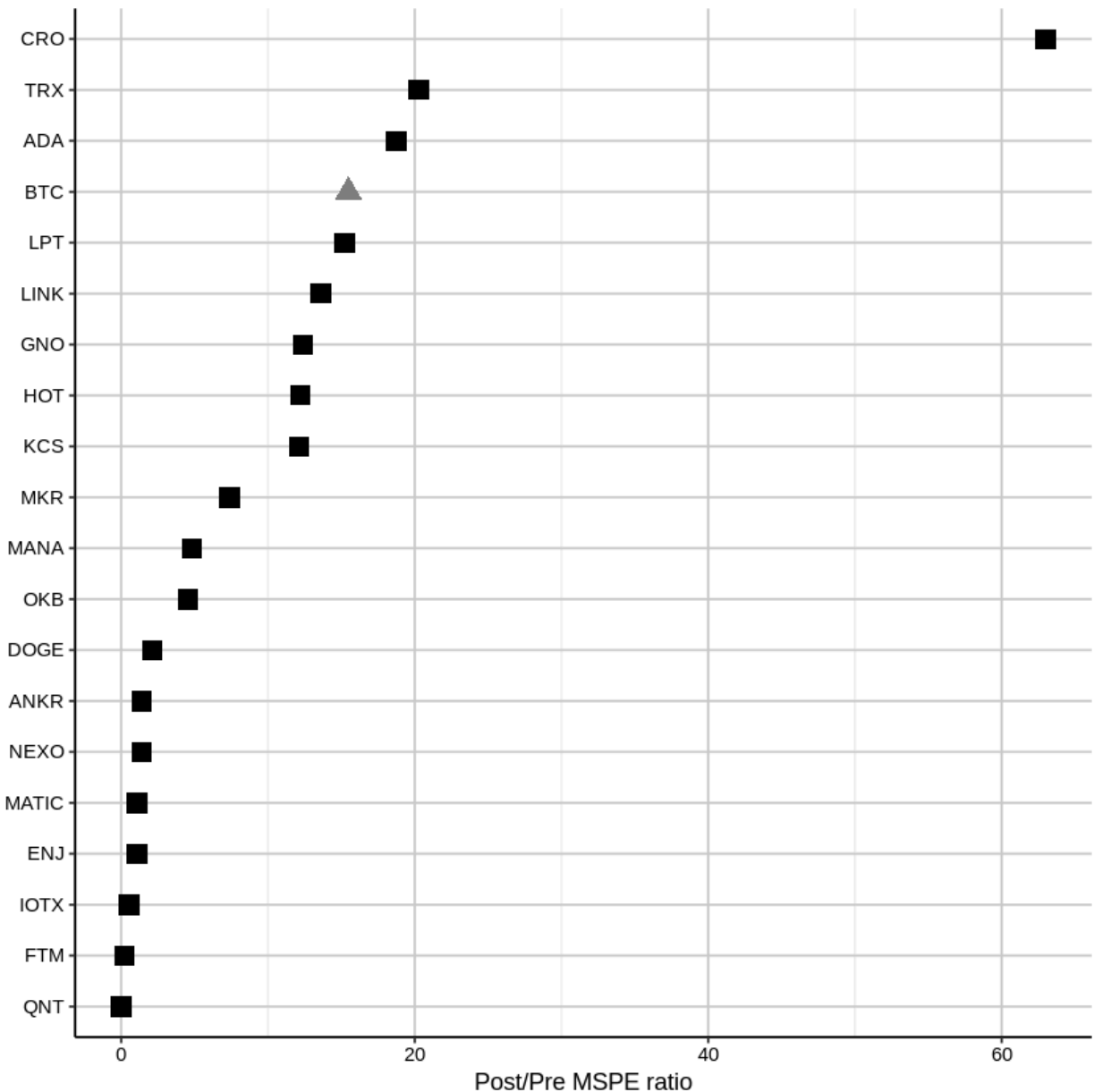

**Figure 16.** The ratio of Post-halving MSPE to pre-crisis MSPE: Bitcoin and Control Cryptocurrencies.

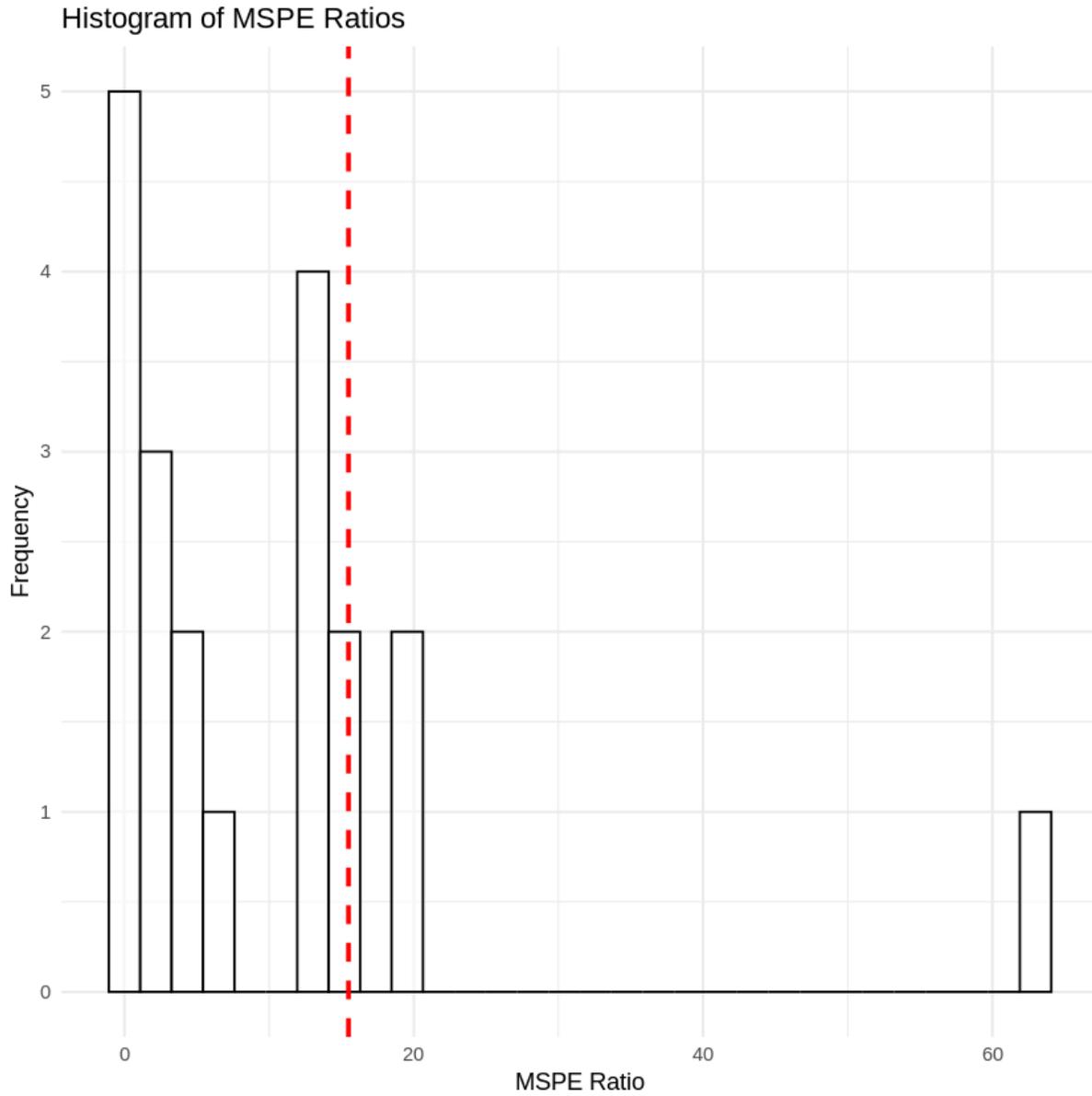

**Figure 16.1.** The histogram shows the ratio of the Post-halving MSPE to pre-crisis MSPE: Bitcoin and Control Cryptocurrencies. The column on the very right corresponds to the ratio of Bitcoin.

*Leave-One-Out Sensitivity Analysis*

This section shows the results of the leave-one-out sensitivity analysis for the original SCM model that excludes each donor that received a weight one at a time. The "leave-one-out cross-validation" (LOOCV) approach in this context means creating several versions of the synthetic control, each time omitting one of the donor units that had a non-zero weight in the original synthetic control. I am using this method to assess the robustness of the results. By iteratively leaving one country out and recalculating the synthetic control, I intend to identify if any single donor cryptocurrency is disproportionately influencing the outcome.

*2024*

The figures below confirm the robustness of the SCM to the exclusion of any non-zero-weighted donor cryptocurrency. Notably, the exclusion of TRX and FET results in a treatment effect almost double that of the original model. However, the weak pre-treatment fit of the model with TRX excluded indicates that this treatment effect is unreliable. The model with FET excluded shows that the divergence begins over two months earlier. Despite these fluctuations, the overall direction of the impact of the 2024 Bitcoin halving stays consistent.

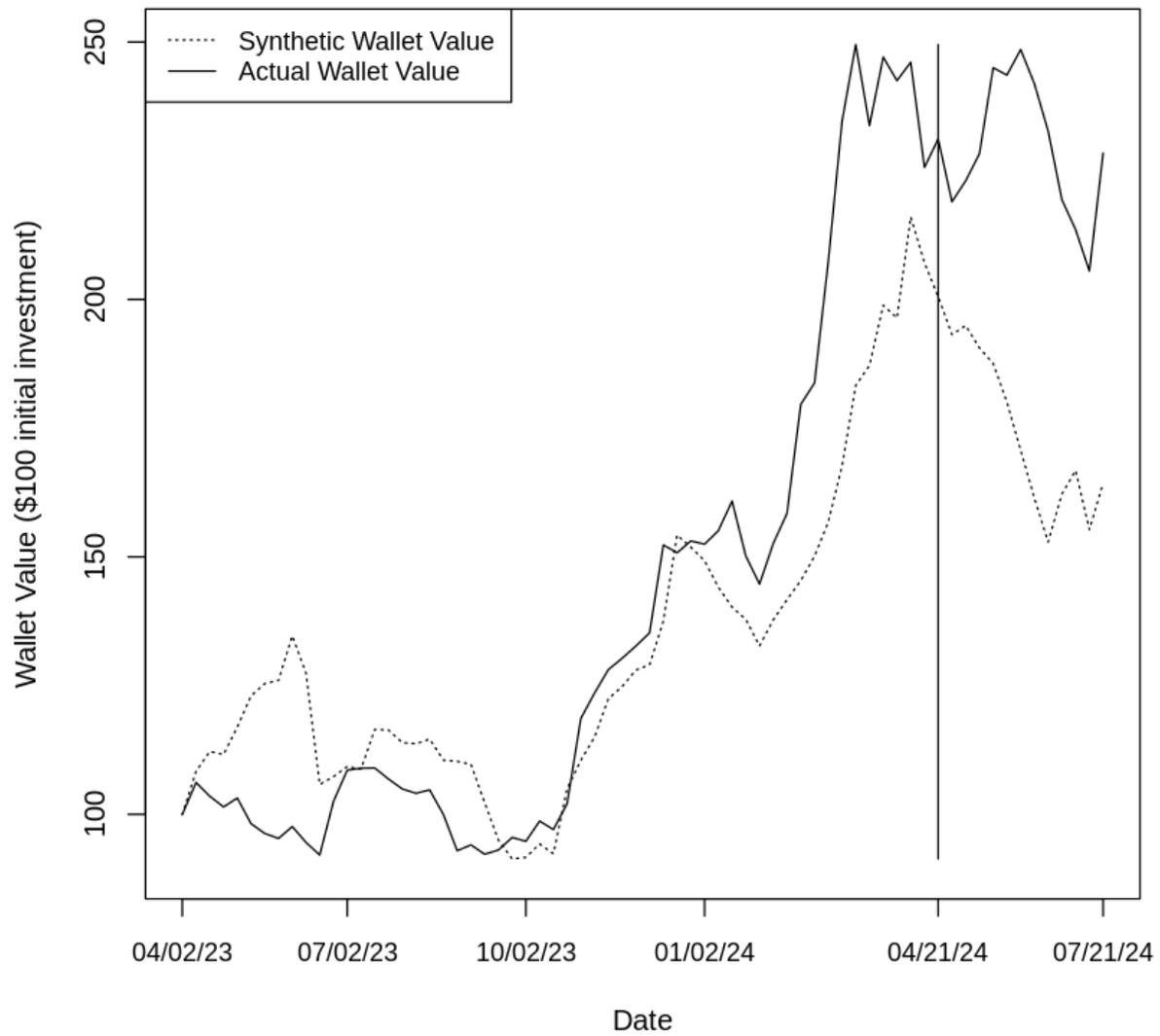

**Figure 17**. This plot shows the trends in the Wallet Value of Bitcoin and its synthetic counterpart (TRX is eliminated). The mean post-treatment gap in this model is 53.8047. The pre-treatment fit is noticeably worse.

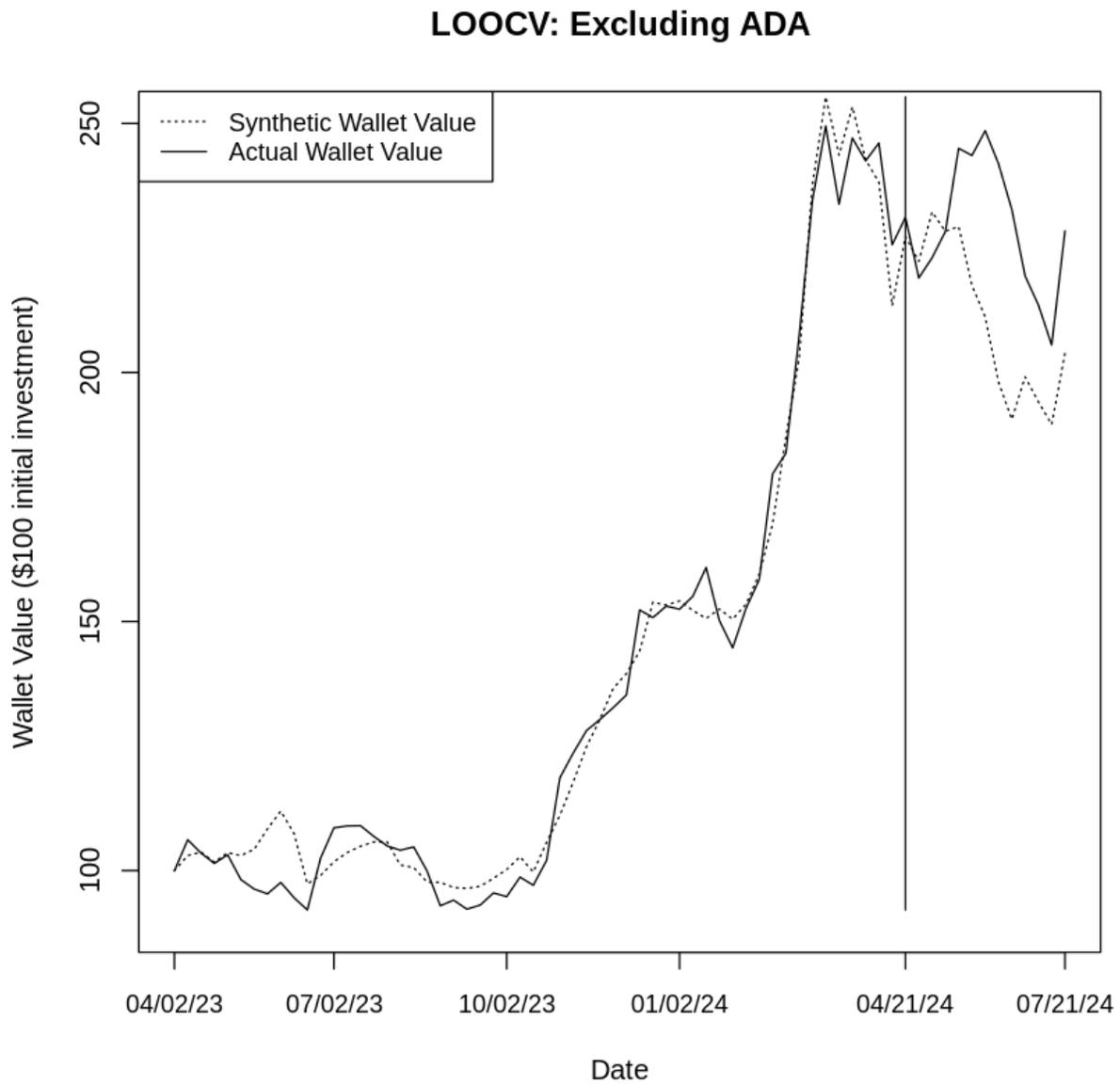

**Figure 18**. This plot shows the trends in the Wallet Value of Bitcoin and its synthetic counterpart (ADA is eliminated). The mean post-treatment gap in this model is 18.1991.

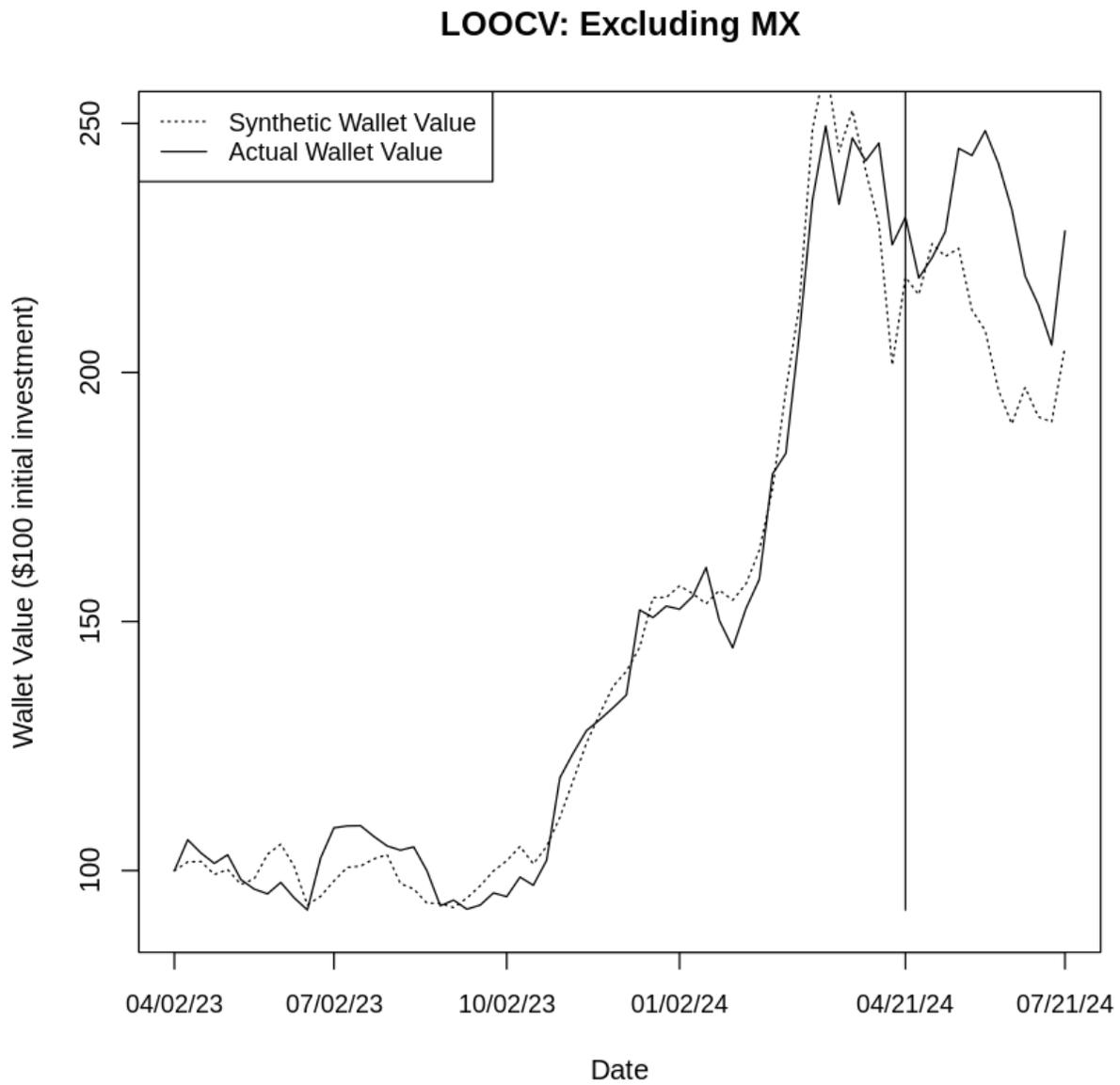

**Figure 19**. This plot shows the trends in the Wallet Value of Bitcoin and its synthetic counterpart (MX is eliminated). The mean post-treatment gap in this model is 21.6026,

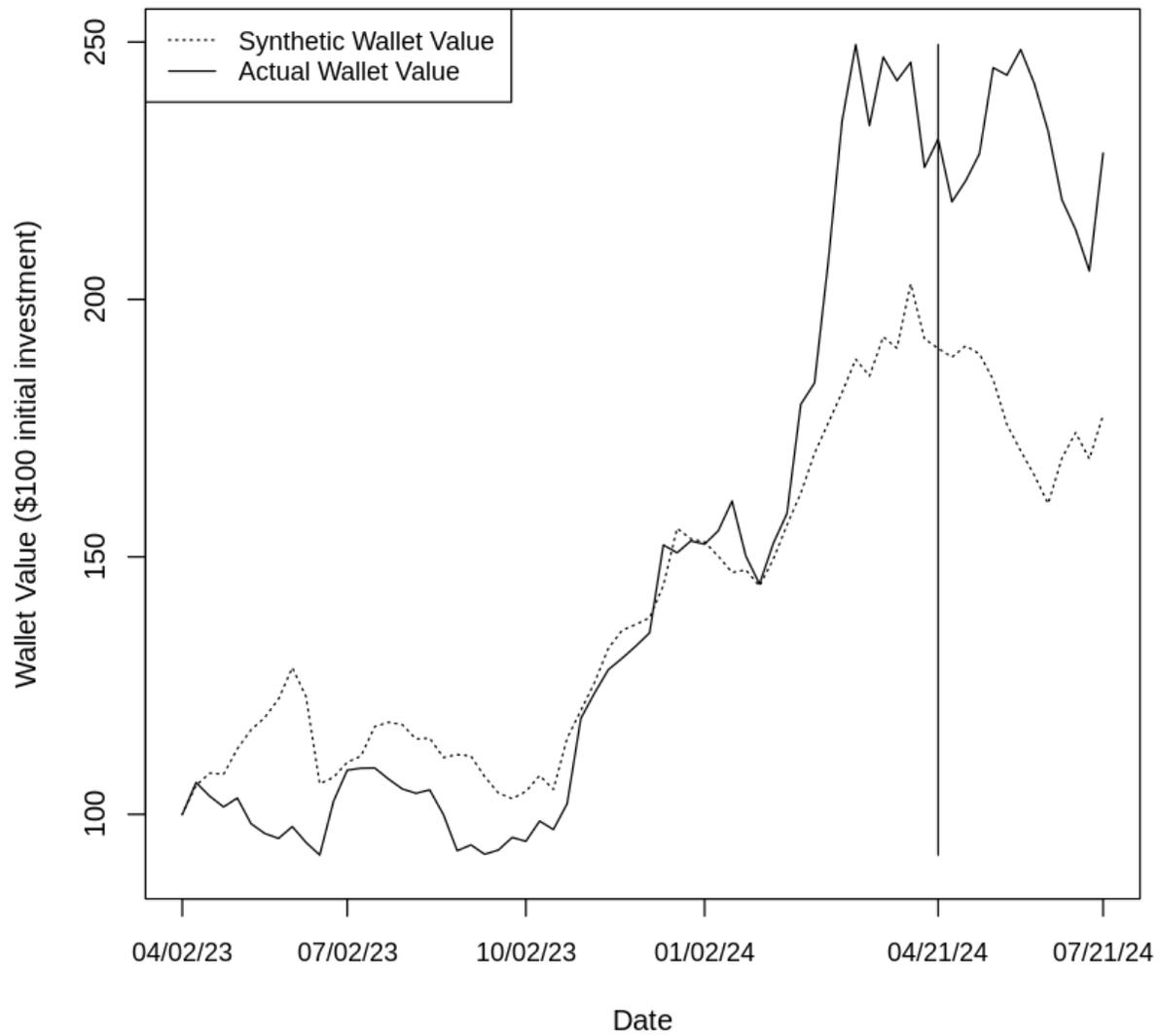

**Figure 20**. This plot shows the trends in the Wallet Value of Bitcoin and its synthetic counterpart (FET is eliminated). The mean post-treatment gap in this model is 51.8007. The divergence begins earlier.

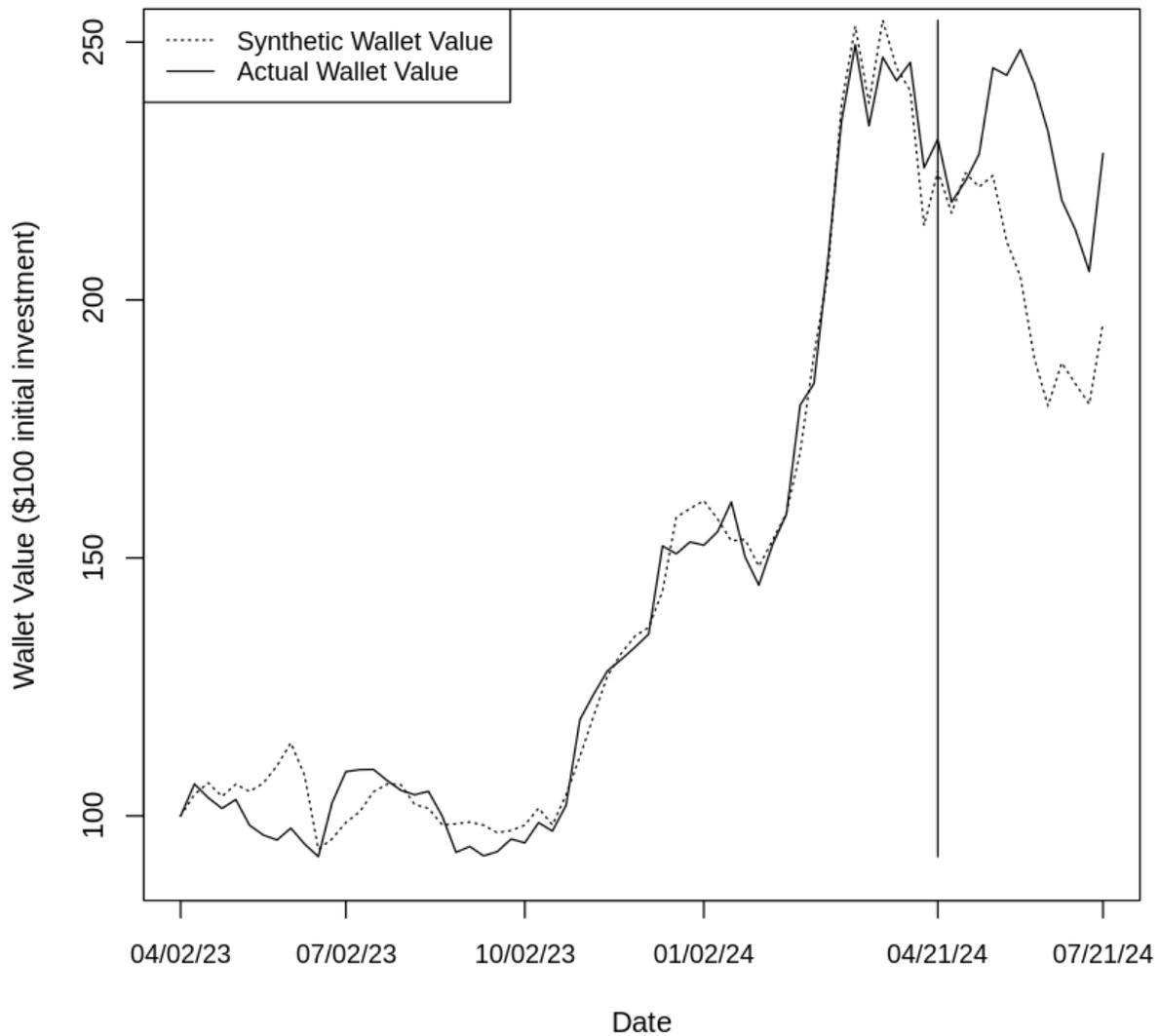

**Figure 21**. This plot shows the trends in the Wallet Value of Bitcoin and its synthetic counterpart (KCS is eliminated). The mean post-treatment gap in this model is 25.8690.

| Leave-One-Out Donor Cryptocurrency (Abbreviation) | Average Wallet Value Gap |
|---|---|
| None (all donors included) | 24.554 |
| <u>TRX</u> | <u>53.805</u> |
| ADA | 18.199 |
| MX | 21.603 |
| <u>FET</u> | <u>51.801</u> |
| KCS | 25.869 |

**Table 7**. Average Wallet Value Effect in the Leave-One-Out sensitivity analysis trials. The most outstanding values are underscored.
*2020*

The figures below fail to confirm the robustness of the SCM to the exclusion of non-zero-weighted donor cryptocurrencies. The exclusion of LINK flips the sign of the effect, implying that LINK exerts a disproportionate influence on the synthetic control. The leave‑one‑out analysis signals that the effect is not robust to minor changes in donor selection. Therefore, one cannot definitively conclude a negative treatment impact of the 2020 Bitcoin halving – the method's assumptions break down when a single donor so strongly influences the counterfactual.

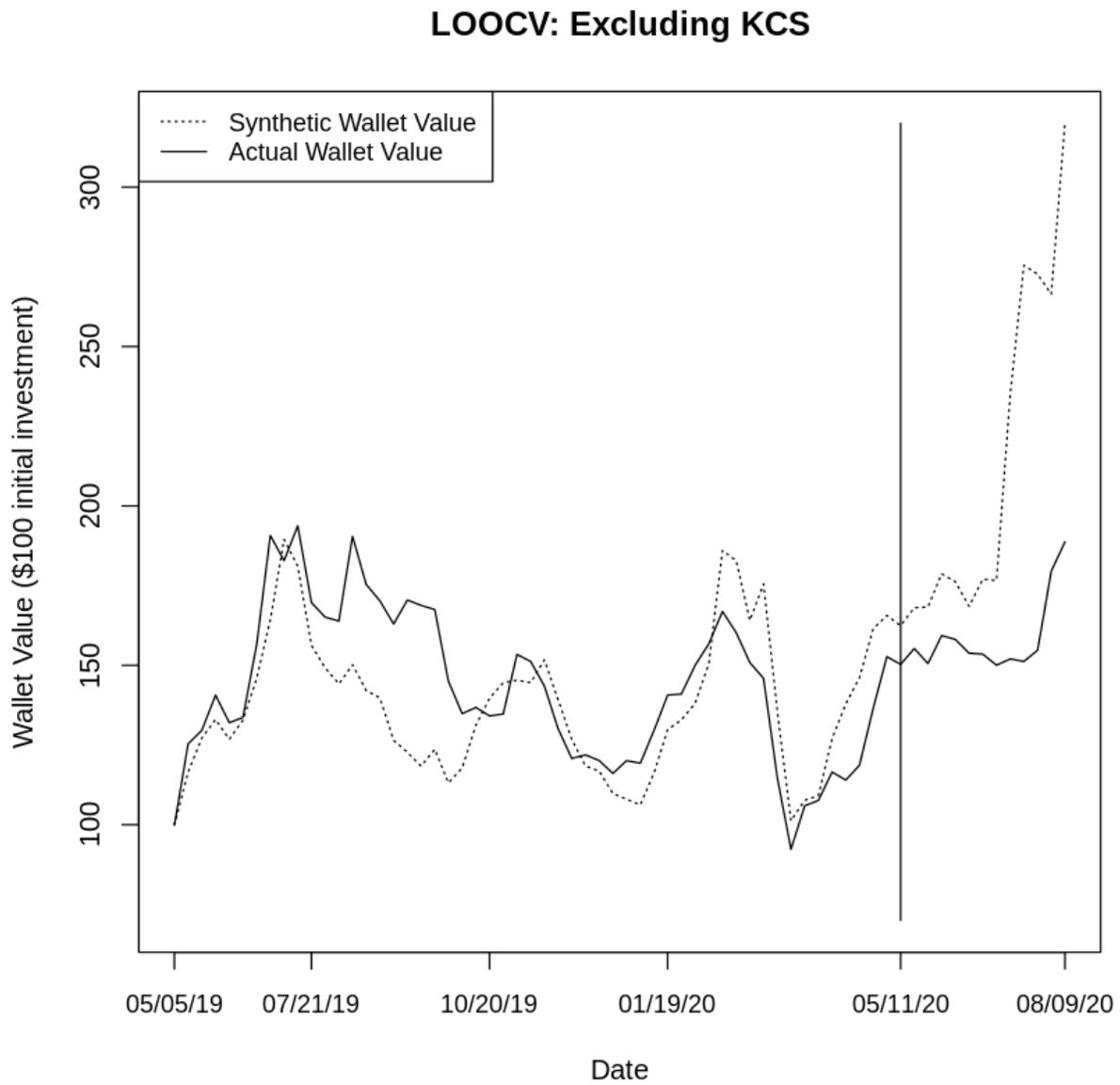

**Figure 22**. This plot shows the trends in the Wallet Value of Bitcoin and its synthetic counterpart (KCS is eliminated). The mean post-treatment gap in this model is -52.9696.

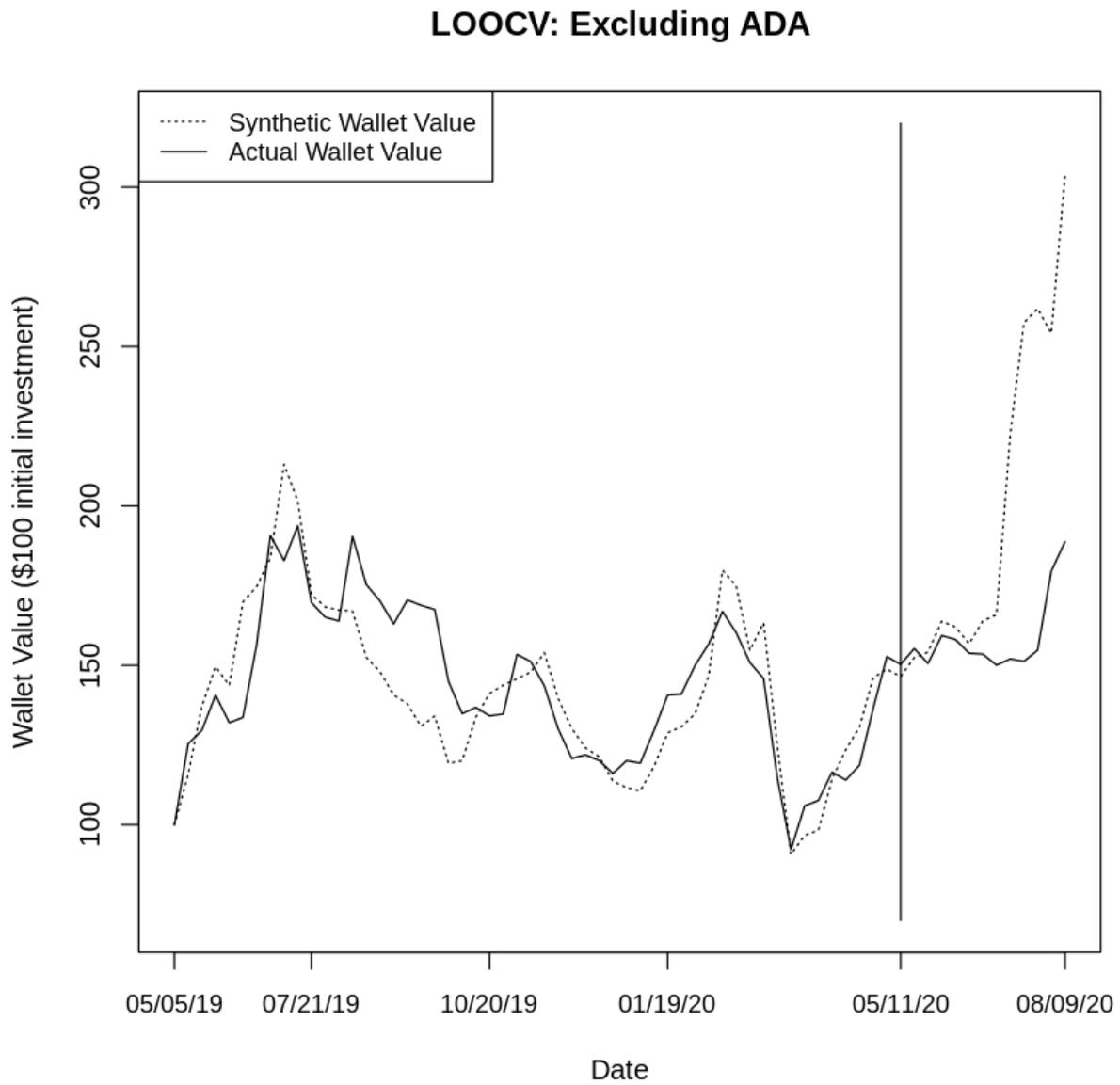

**Figure 23**. This plot shows the trends in the Wallet Value of Bitcoin and its synthetic counterpart (ADA is eliminated). The mean post-treatment gap in this model is -39.0453.

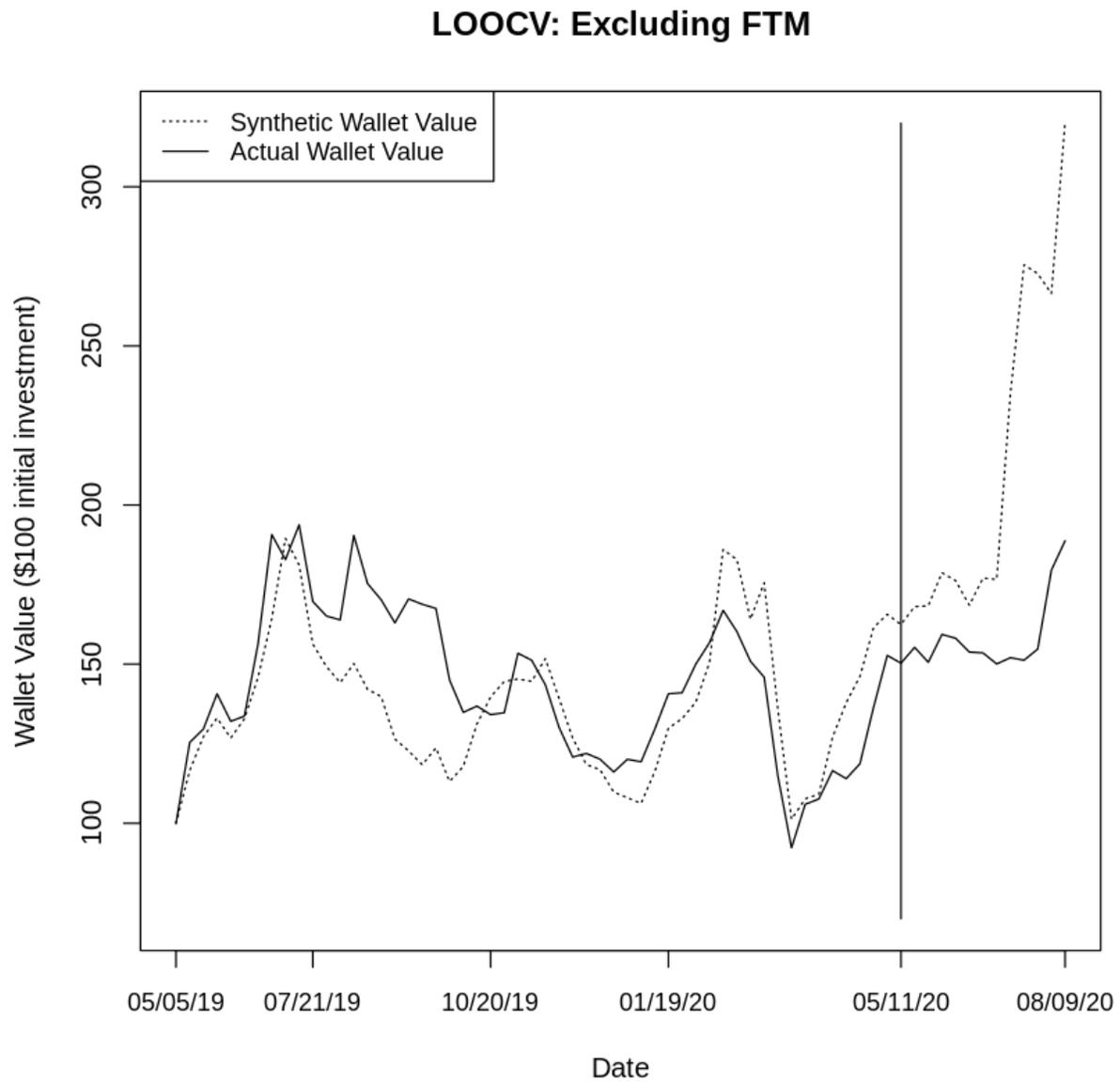

**Figure 24**. This plot shows the trends in the Wallet Value of Bitcoin and its synthetic counterpart (FTM is eliminated). The mean post-treatment gap in this model is -52.9742.

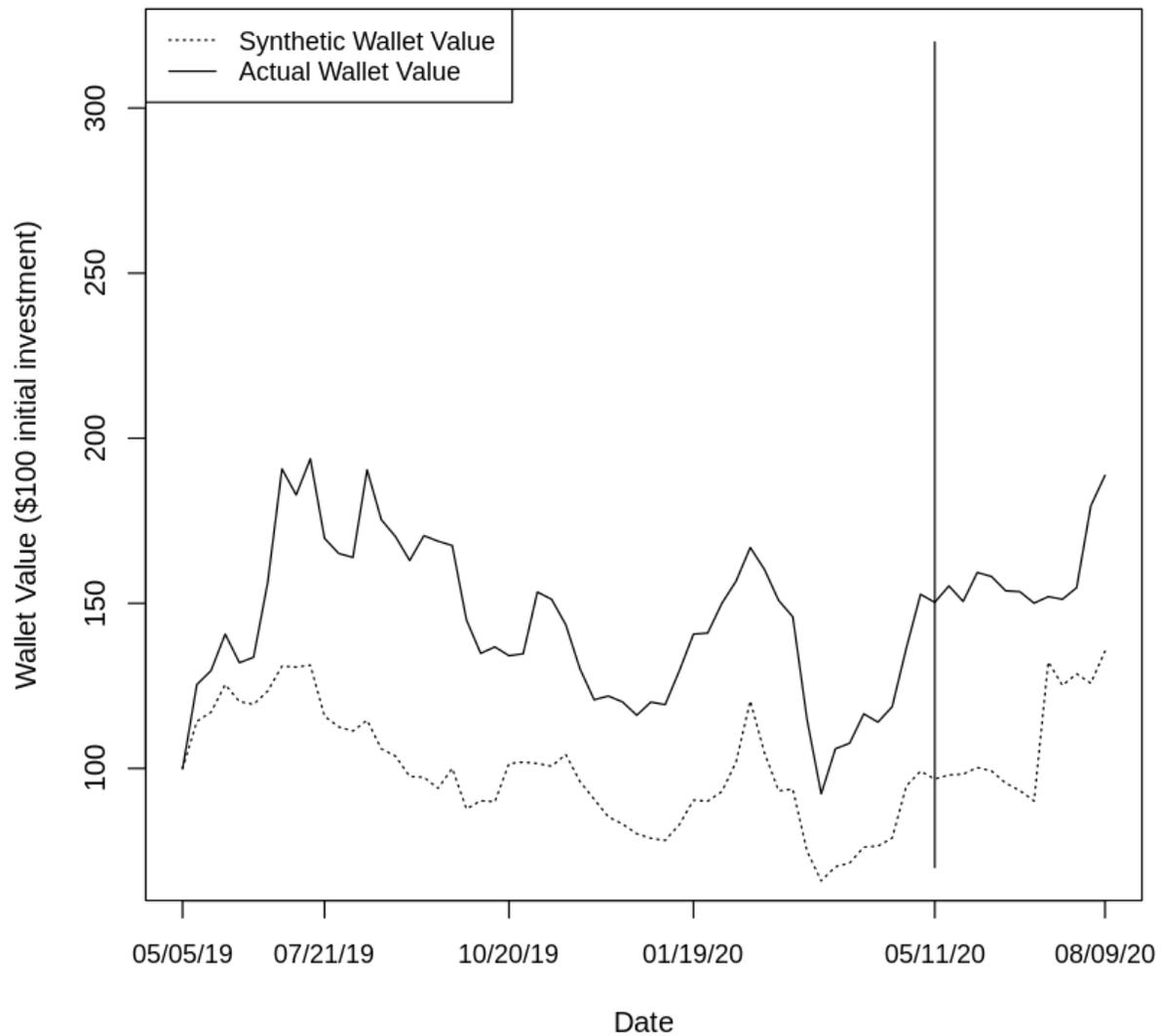

**Figure 25**. This plot shows the trends in the Wallet Value of Bitcoin and its synthetic counterpart (LINK is eliminated). The mean post-treatment gap in this model is 49.1011. The pre-treatment fit is almost absent.

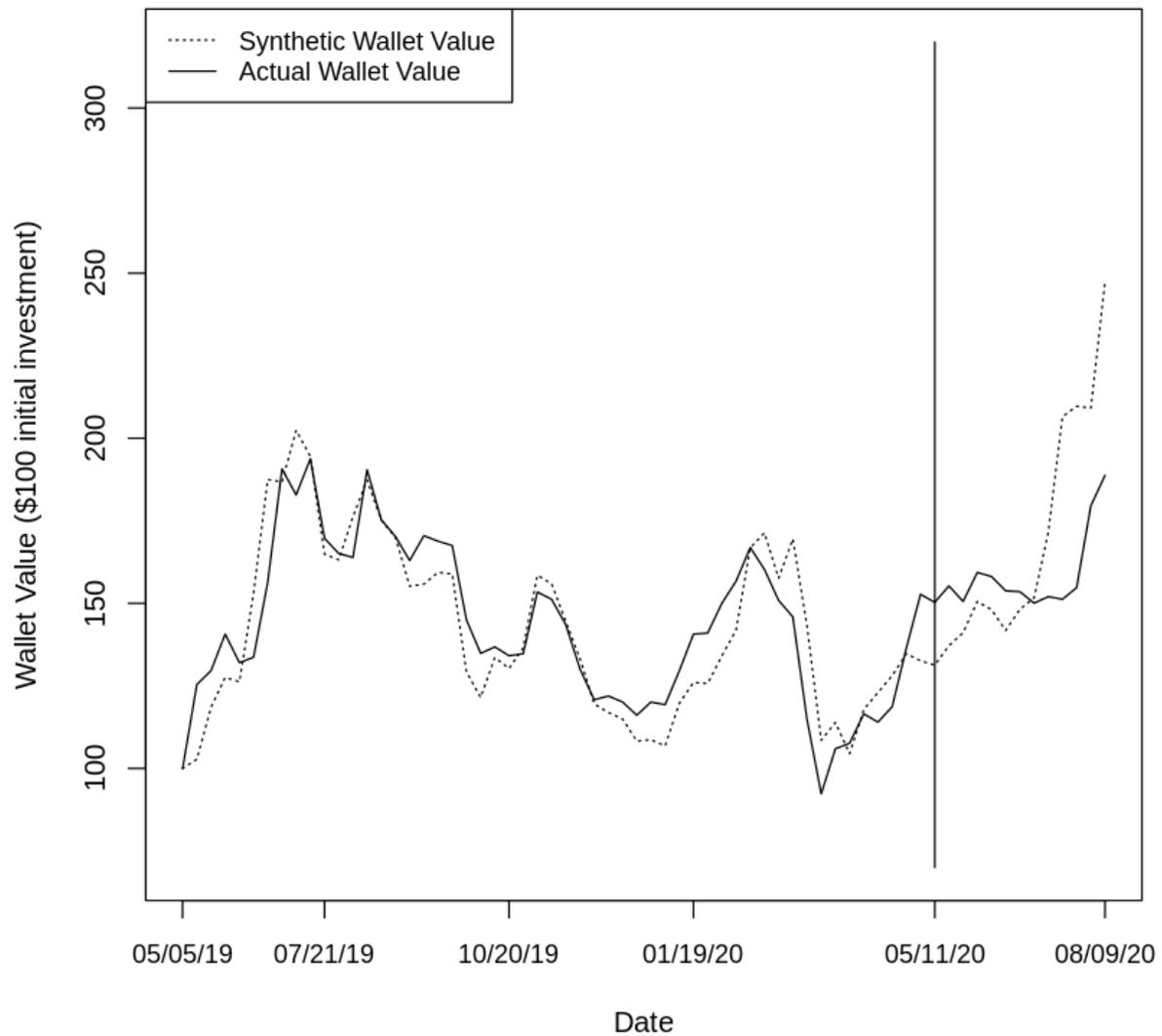

**Figure 26**. This plot shows the trends in the Wallet Value of Bitcoin and its synthetic counterpart (DOGE is eliminated). The mean post-treatment gap in this model is -10.5205.

| Leave-One-Out Donor Cryptocurrency (Abbreviation) | Average Wallet Value Gap |
|---|---|
| None (all donors included) | -29.116 |
| KCS | -52.970 |
| ADA | -39.045 |
| FTM | -52.974 |
| <u>LINK</u> | <u>49.101</u> |
| DOGE | -10.521 |

**Table 8**. Average Wallet Value Effect in the Leave-One-Out sensitivity analysis trials. The most outstanding value is underscored, showing that the SCM is especially sensitive to the removal of LINK from the donor pool.

**Summary and Conclusion**

A multitude of academic and non-academic publications have linked the Bitcoin halving to subsequent Bitcoin price increases. However, to date, no rigorous causal inference research has been conducted to confirm this causal link. This 2024 halving study presents robust evidence of a positive impact of the 2024 Bitcoin halving on its price three months later. I demonstrate a 24.55% average gap between Bitcoin's "Wallet Value" (a proxy of the price I derive for comparability) and the "Wallet Value" of a comparable synthetic Bitcoin without halving. The gap emerges over a period of twelve months, which is approximately one-fifth of the total percentage change in the price of Bitcoin during the study period – from April 2nd, 2023, to July 21st, 2024 (17 months). This is an encouraging piece of evidence suggesting that despite being so well-anticipated, Bitcoin halving is not priced in, and market reactions to it still occur. The 2020 halving study fails to obtain a statistically significant and robust causal estimate of the effect of the 2020 Bitcoin halving on Bitcoin's price. This is expected given the broader macroeconomic dynamics that influenced the cryptocurrency market during that period, making it hard for the Synthetic Control Model to pick up the intended treatment.

Further research can build on this work in multiple ways. First, it would be interesting to build similar models to study the effects of all the previous Bitcoin halvings starting from 2012 using monthly panel data instead of weekly to further smooth volatility, thus minimizing the risk of overfitting due to random noise. More outcome variables apart from price or its proxy could be considered – I show a few specific examples in the "Y-variable placebo" section in Appendix 5. This could help uncover patterns in the longer-term effect and the consistency (or lack thereof) of its magnitude. Furthermore, a more comprehensive set of predictors with strong predictive power (Abadie, 2021) without missing data should be prioritized since the "dataprep" function of the SCM cannot handle missing values. Second, a similar setup of the SCM could be used to estimate the causal effect of other less anticipated events in the cryptocurrency world, like the approvals and launches of exchange-traded funds (ETFs), flash crashes, market manipulations, exchange insolvencies, bank runs, black swan events, hashrate shocks, wallet movements, or hard forks. Finally, a more robust approach be employed to identify idiosyncratic shocks that would likely not have affected the outcome variable for donor units if the treatment did not take place could – for example, using new "deep research" tools offered by leading AI companies.

## Appendix 1: Obtaining Source Data from IntoTheBlock

Here is how I obtained and transformed all the treatment and covariate data on September 24, 2024:

**Step 1**: Launch IntoTheBlock Web App

**Step 2**: Sign up / Log in using a Pro Plan with Unlimited CSV Data Downloads (Institutional Plan with IntoTheBlock REST API will work as well)

**Step 3**: Download all available data from the "Financial," "Network," and "Ownership" categories.

**Step 4**: Run the following R scripts for ADA, ALGO, ANKR, BTC, CRO, DOGE, ENJ, ETH, FET, FTM, GNO, HOT, IOTX, JET, KCS, LEO, LINK, LPT, MANA, MATIC/POL, MKR, MX, NEXO, OKB, QNT, TON, and TRX:

- Cleaning Step 1.R for merging all cryptocurrency data on the 'DateTime' column, one file per cryptocurrency;

- Cleaning Step 2.R for standardizing column names;

- Cleaning Step 3.R for converting the daily data to weekly;

- Cleaning Step 4.R for merging all cryptocurrency files into one long .csv.

## Appendix 2: Complete List of Variables

| Category | Variable | Meaning |
|---|---|---|
| Financial | Historical In/Out of the Money | For any address with a balance of tokens, ITB identifies the average price (cost) at which those tokens were purchased and compares it with the current price. If Current Price > Average Cost, the address is "In the Money." If Current Price < Average Cost, the address is "Out of the Money." |

| Financial | Historical Active Addresses by Profitability | Daily active blockchain addresses over time categorized based on their profitability status: in the money (profiting), at the money (breaking even) or out of the money (losing money). |
|---|---|---|
| Financial | Historical Break Even Price | For any given day, the addresses that had historically made a profit vs. the addresses that had made a loss until that day and were still holding the asset |
| Financial | Number of Large Transactions | Daily number of on-chain transactions greater than $100,000. |
| Financial | Large Transactions Volume | Aggregated daily volume, measured in onchain transactions where each transaction was greater than $100,000. The adjusted view subtracts the change in UTXO (unspent transaction outputs) based transactions. |
| Financial | Large Transactions Volume in USD | Aggregated daily volume, measured in USD from onchain transactions where each transaction was greater than $100,000. The adjusted view subtracts the change in UTXO-based transactions. |
| Financial | Bulls And Bears | Bull: An address that bought more than 1% of the volume traded in the last 24 hours. Bear: An address that sold more than 1% of the volume traded in the last 24 hours. |
| Financial | Price | Price by day (USD). |
| Financial | Market Cap | The total value of all the circulating supply of a cryptocurrency. It's calculated by multiplying the current market price by the total number of coins that are currently in circulation. |
| Financial | Market Value to Realized Value Ratio | Ratio of Market Cap to the average purchasing cost of each address. |
| Financial | Network Value to Transaction (NVT) Ratio | Ratio of market cap to transacted value in dollars. |
| Financial | Asset Market Cap Comparison | It shows the historical results of dividing the market capitalization of the asset by the market capitalization of BTC and ETH. |
| Financial | Average Transactions Size | Total value of transactions divided by the number of transactions. |
| Financial | Average Balance ($) | Market cap divided by number of addresses (with a balance of |

| | | tokens). |
|---|---|---|
| Financial | Volatility | Annualized Price volatility using 365 days |
| Financial | Beta Coefficient | This indicator evaluates a cryptocurrency's volatility relative to Bitcoin's, offering insights into risk and price fluctuations compared to a benchmark. |
| Financial | Historical Correlations | An historical comparison of the values for the 30-day correlation coefficient between the crypto asset being monitored and a basket of the other most valuable ones. |
| Financial | Circulating Supply | Supply in circulation. |
| Network | Total Fees | Total amount paid in fees to use the blockchain on a given day. |
| Network | Average Transaction Fee | Daily mean transaction cost. |
| Network | Daily Active Addresses | New Addresses: New addresses created. Zero Balance Addresses: Addresses that transferred out all of their tokens. Active Addresses: Addresses that made a transaction. |
| Network | Total Addresses | Subtracts those addresses that have no balance ("Total Zero Balance") from the total number of addresses in the network ("Total") to arrive at those addresses that actually have a balance ("Total With Balance"). |
| Network | Active Addresses Ratio | The Active Addresses Ratio is calculated as the ratio of Active Addresses to Addresses with a Balance, expressed as a percentage. Note: In some cases, especially for coins with few holders, this ratio can exceed 100%. This can happen when all the balance from one address is transferred to a new address. In such cases, the number of "Active Addresses" would be 2, while the number of "Addresses with a Balance" would remain 1, leading to a ratio greater than 100%. |
| Network | Activity to Fee Ratio | This ratio of Daily Active Addresses to Blockchain Fees. This offers insights into network health, usage value, and the potential impact of high fees. |
| Network | New Adoption Rate | The percentage of new addresses making their first transaction out of all active addresses on a given day. This provides insight into the share that newcomers make out of total activity. |
| Network | Address Birth-Death | The ratio of new addresses being created to addresses with a |

| | Ratio | balance that have not made a transaction in over one year. This indicator contrasts 'New Addresses' with inactive ones over a year, offering insights into the growth and recurring activity of a blockchain's user base. |
|---|---|---|
| Network | Number of Transactions | We count as a number of transactions those ones that are valid. This means that those transactions that fail and are reverted are not included in this metric. |
| Network | Transaction Count by Size | Amount (or percentage) of transactions based on a given amount. |
| Network | Transactions Volume | On-chain volume transferred in the period. |
| Network | Transactions Volume in USD | On-chain volume transferred in the period measured in USD given daily price. |
| Network | Transaction Volume in USD by Size | Amount (or percentage) of transaction volume based on a given amount. |
| Network | Average Time Between Transactions | The percentage of Addresses With a Balance that had a transaction during a given period (Active Addresses / Addresses with a Balance). |
| Network | Holding Time Of Transacted Coins | A measurement of how much time the coins have been held before being transacted. |
| Network | Created Unspent Transaction Outputs | Count/volume of UTXO by day plotted against the price by day in USD. |
| Network | Spent Unspent Transaction Outputs | Count/volume of UTXO that was spent by day plotted against the price by day in USD. |
| Network | Unspent Transaction Outputs Age | Volume of UTXO by age (UTXO Age distribution). |
| Ownership | Historical Concentration | The Concentration group of indicators shows the allocation and activity of a crypto asset's circulating supply based on three groups: whales, investors, and retail. |
| Ownership | Large Holders Inflow | Variation over time of the inflows of large holders (those possessing at least 0.1% of the circulating supply). |
| Ownership | Large Holders Outflow | Variation over time of the outflows of large holders (those possessing at least 0.1% of the circulating supply). |
| Ownership | Large Holders Netflow | Variation over time of the netflows of large holders (those |
| | | possessing at least 0.1% of the circulating supply). |
| Ownership | Addresses by Time Held | The variation over time of the number of addresses that have been holding some coins. These addresses are divided between Hodlers, Cruisers, and Traders, depending on how much time they have been holding their coins. |
| Ownership | Balance by Time Held | The variation over time of the balance of coins that each group has been holding. These groups are divided into Hodlers, Cruisers, and Traders, depending on how much time they have been holding their coins. |
| Ownership | All-Time Highers / Lowers | All-Time Higher: An address (with a balance) that bought within 20% of the token's all-time high price. All-Time Lowers: An address (with a balance) that bought within 20% of the token's all-time low price. |
| Ownership | Addresses by Holdings | Number of addresses holding the selected crypto/dollar amount. This can be displayed in terms of the aggregate count of addresses or the percentage they make out of all. |
| Ownership | Balance by Holdings | The total volume in crypto/dollar terms held by the selected group of addresses. This can be displayed in terms of the aggregate amount of crypto held or the percentage they make out of the circulating supply. |
| Exchanges | Inflow Volume | Total amount (in $ or tokens) entering exchange(s) deposit wallets. "All Exchanges" refers to all supported exchanges. |
| Exchanges | Inflow Transaction Count | Total number of deposit transactions. "Aggregated Exchanges" refers to all supported exchanges. |
| Exchanges | Outflow Volume | Total amount (in $ or tokens) leaving exchange(s) withdrawal wallets. "Aggregated Exchanges" refers to all supported exchanges. |
| Exchanges | Outflow Transaction Count | Total number of withdrawal transactions. "Aggregated Exchanges" refers to all supported exchanges. |
| Exchanges | Net Flows | Netflows show the difference between tokens entering an exchange minus those leaving exchanges. "Aggregated Exchanges" refers to all supported exchanges. The change shows the net increase/decrease in exchanges' holdings over the respective time. |
| Exchanges | Large Holders Netflow to Exchange | The ratio of large holders' netflows (inflows - outflows) over |

| | Netflow Ratio | centralized exchanges netflows. This indicator, by comparing 'Exchange Netflows' and 'Large Holders Netflows', provides insights into retail and large holder activity, signaling possible market trends like cashing out or asset accumulation. |
|---|---|---|
| Exchanges | Total Flows | Sum of the amount entering an exchange plus the amount leaving an exchange. This is an indicator of overall exchange activity. "Aggregated Exchanges" refers to all supported exchanges. |
| Order Books | Bid-Ask Spread | A measure of liquidity calculating the difference between the asking price and the bid price in a limit order book. This indicator refreshes every 1 minute while the last 12 hours of activity are displayed. |
| Order Books | Trades per Side | Measures the number (or volume) of trades where the buyers "crossed the spread" and bought at the Ask price vs the number (or volume) of trades where sellers "crossed the spread" and sold at the Bid price per minute. The last 12 hours of activity are displayed in the graph. |
| Social | Telegram Members | Uses a classification machine learning technique to determine if the texts used in the Telegram messages related to a given token have a positive, negative, or neutral connotation. |
| Social | Telegram Sentiment | Number of members in the token's official Group for a given period. |
| Social | GitHub | Github indicators provide insights regarding development activity for a crypto asset based on commits (changes made to the code of the asset ecosystem by developers), stars and issues (interest and engagement shown by the community), and pull requests (changes and network improvements submitted and approved over time). |
| Derivatives: Perpetuals | Price | Price of a perpetual swaps contract for a given exchange. |
| Derivatives: Perpetuals | Funding Rate | The fee charged to perpetual swap holders to keep prices of the contract pegged to the underlying spot price. The funding rate is a function of the contract's premium (discount) and the exchange's interest rate. If the funding rate is positive, long holders have to pay that percentage for each contract they hold to the short holders, and vice versa. The funding rate is typically distributed every 8 hours and can act as a short-term indicator of market positioning. |
| Derivatives: Perpetuals | Volume | Dollar amount traded in a 24-hour period. Average daily volume is the average total volume traded daily over the last seven days. The price is the weighted average of all contracts' prices weighted by their open interest. |
| Derivatives: Perpetuals | Open Interest | Dollar amount of contracts outstanding (open positions). Average open interest refers to the average open interest on a given day taking into account data for the last seven days. The price is the weighted average of all contracts' prices weighted by their open interest. |
| Derivatives: Perpetuals | Open Interest to Market Cap Ratio | Percentage that perpetual swaps' open interest makes up out of the asset's market cap. |
| Derivatives: Perpetuals | Turnover Ratio | Ratio of the 24-hour volume over open interest. This ratio indicates the level of short-term hedging and speculative activity for a derivatives contract relative to its existing positions. Volatility shown is 30-day price volatility. |
| Derivatives: Perpetuals | Basis | Percentage difference of the contract price minus the spot price |

**Table 9**. Complete List of Variables Available in the Sourced Dataset. Most variables were not used due to a large share of NAs.

## Appendix 3: Potential Donors

*2024 Halving*

2    ADA

3    ALGO

4    ANKR

5    CRO

7    ENJ

8    ETH

9    FET

10    FTM

| 11 | GNO |
| 12 | HOT |
| 13 | IOTX |
| 15 | KCS |
| 16 | LEO |
| 17 | LINK |
| 18 | LPT |
| 19 | MANA |
| 20 | MATIC |
| 21 | MKR |
| 22 | MX |
| 23 | NEXO |
| 24 | OKB |
| 25 | QNT |
| 26 | TRX |

*2020 Halving*

| 2 | ADA |
| 4 | ANKR |
| 5 | CRO |
| 6 | DOGE |
| 7 | ENJ |
| 8 | ETH |
| 10 | FTM |
| 11 | GNO |
| 12 | HOT |
| 13 | IOTX |
| 15 | KCS |
| 17 | LINK |
| 18 | LPT |
| 19 | MANA |
| 20 | MATIC |
| 21 | MKR |
| 23 | NEXO |
| 24 | OKB |
| 25 | QNT |
| 26 | TRX |

**Appendix 4: R Code**

Consolidated R code: https://github.com/virtonen/btchalvingsynth

**Appendix 5: Y-Variable Placebo**

*2024*

I also conduct a different kind of placebo test, choosing predictors like the Share of Hodler Balance Addresses and the Log of the Number of All-Time Highers as the outcome variables instead of the wallet value. In theory, most predictors should not be directly impacted by Bitcoin halving. For this placebo test, the SCM obtains a decent pre-treatment match only for two other variables – the Share of Hodler Balance Addresses and the Log of the Number of All-Time Highers.

Interestingly, the effect of the 2024 halving on the log of the aggregated balance of coins held by addresses classified as "Hodlers" (those holding coins for over one year) is positive. That means that long-term holders accumulated or retained more Bitcoin compared to the synthetic control scenario as a result of the 2024 Bitcoin halving (which might reflect increased confidence or bullish expectations among long-term investors following the halving, consistent with the common idea that halvings encourage holders to anticipate future price increases) (Figure 27).

I uncover a similarly positive effect on the log of the number of all-time highers (the count of addresses that bought within 20% of the cryptocurrency's all-time high price), signifying new adoption around the time of the halving. That makes sense because the popularity of Bitcoin is expected to go up around the time of halving in general, and the phenomenon serves as a reminder of the increasing rate at which Bitcoin moves toward scarcity for investors (Figure 28), potentially nudging investors to buy Bitcoin despite its price being at its all-time highs.

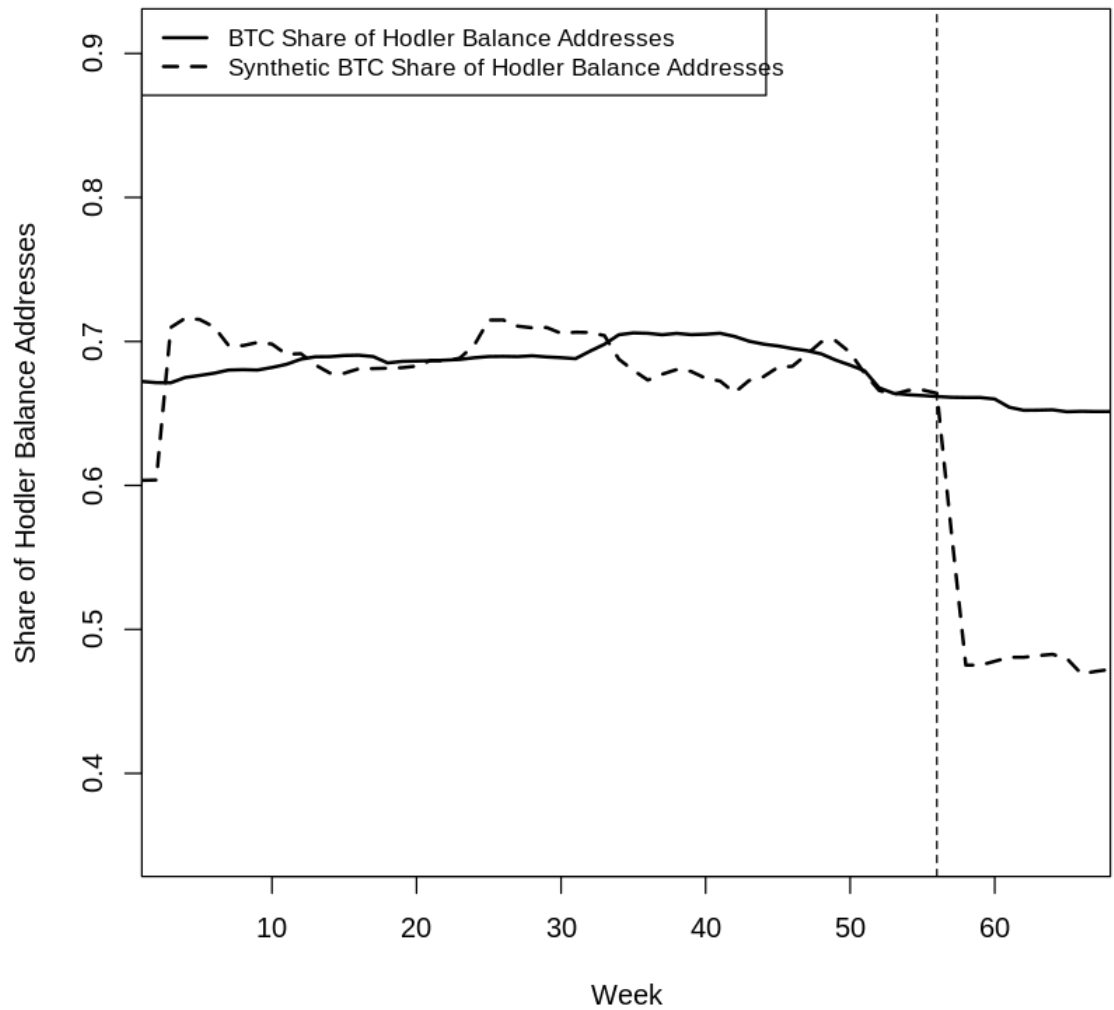

**Figure 27.** Y trend from April 2nd, 2023 (denoted as week 1) until September 9th, 2024 (week 78). The Y variable in this model is the Share of Hodler Balance Addresses. The vertical dotted line shows the time of the 2024 Bitcoin halving.

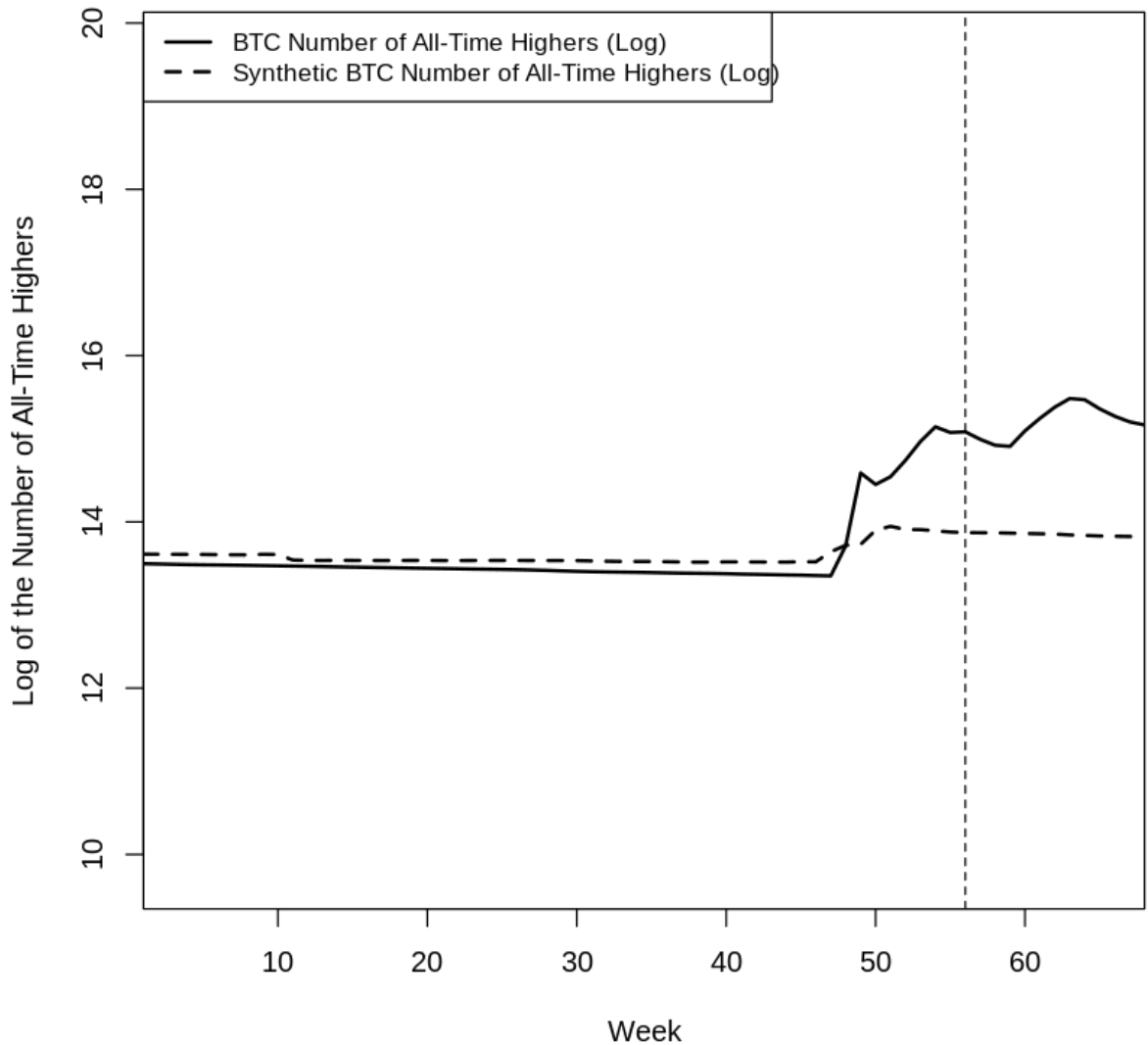

**Figure 28.** Y trend from April 2nd, 2023 (denoted as week 1) until September 9th, 2024 (week 78). The Y variable in this model is the Log of the Number of All-Time Highers. The vertical dotted line shows the time of the 2024 Bitcoin halving.

*2020*

Next, I choose the log of the number of transactions (the count of the number of valid transactions, excluding failed and reverted ones) as the outcome variable instead of the wallet value. This placebo test can be considered passed, as the pre-treatment match is fairly good, and the post-treatment divergence does not occur. All the other Y-variable placebo trials are reported in Appendix 4.

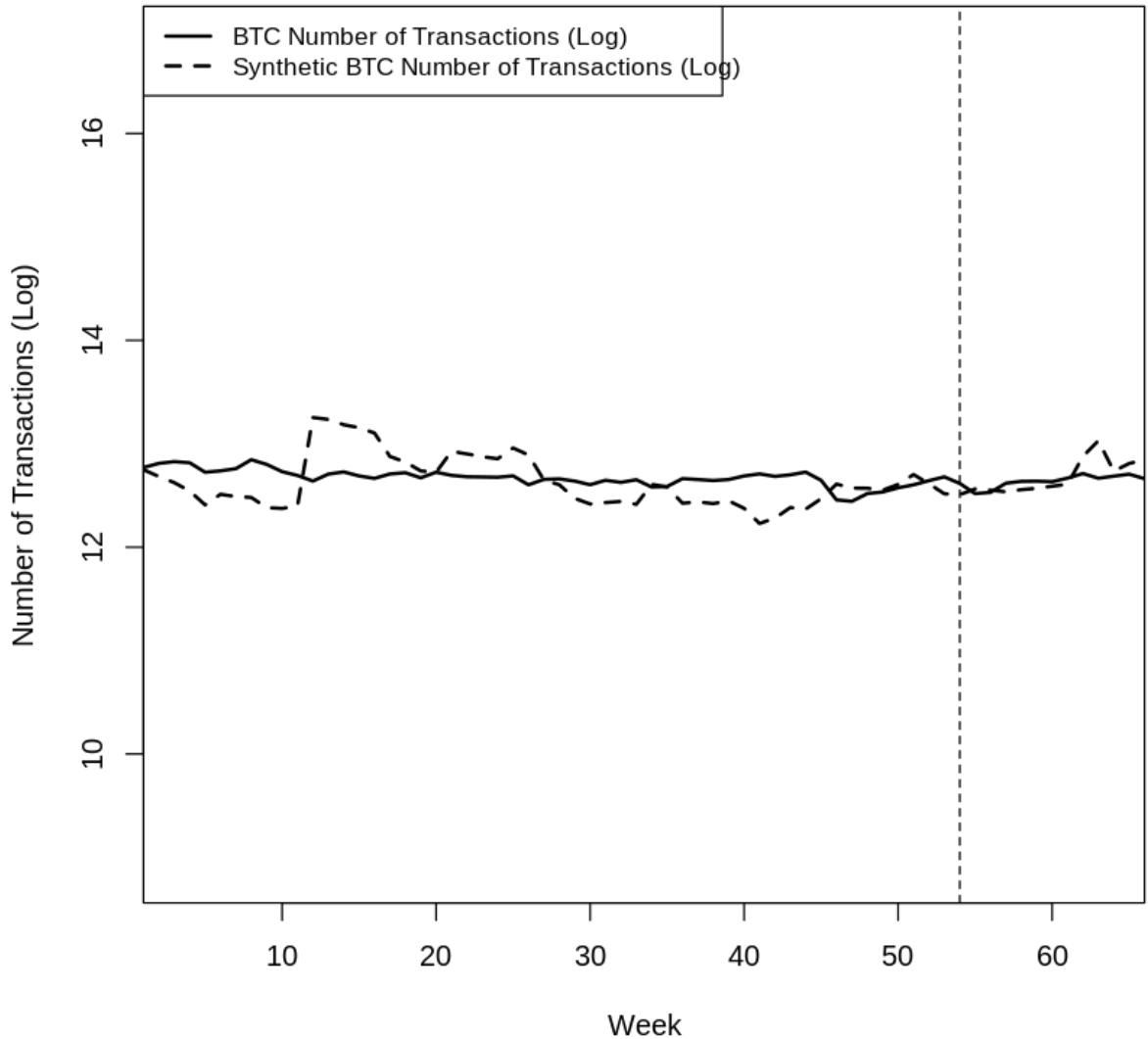

**Figure 29.** Y trend from May 5th, 2019 (denoted as week 1) until June 13th, 2021 (week 110). The Y variable in this model is the log of the number of transactions. The vertical dotted line shows the time of the 2020 Bitcoin halving.

**Appendix 6: Placebo Plots and MSPE Histograms**

*2024*

Placebo study with no MSPE limit:

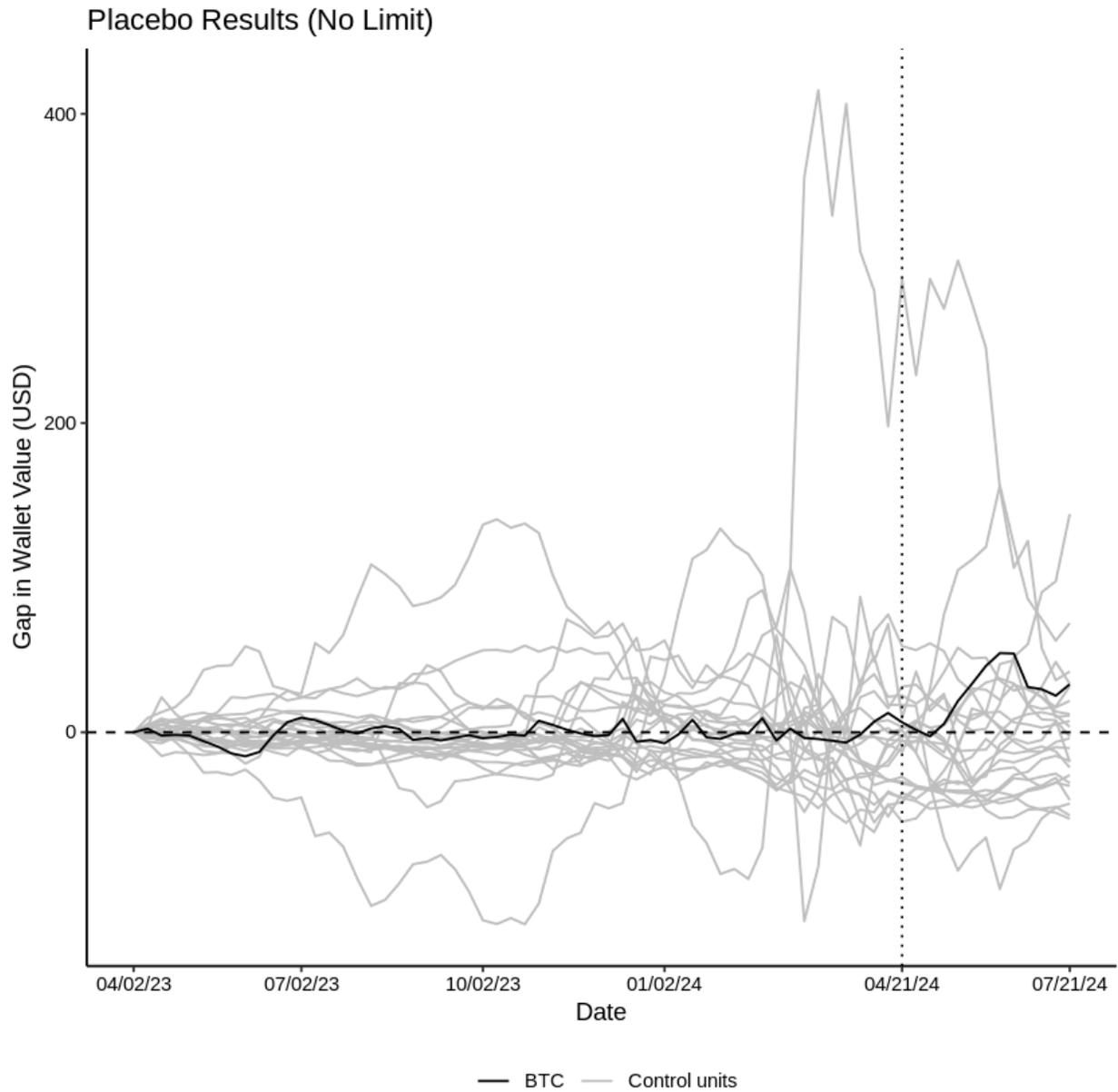

**Figure 30**. This plot shows the difference between observed units and 23 synthetic controls for the treated and control units, in black and grey, respectively. No cryptocurrencies are discarded. Wallet value gaps for Bitcoin and the placebo gaps for control cryptocurrencies are shown. The vertical dotted line shows the time of the 2024 Bitcoin halving.

Placebo study with 100x MSPE limit:

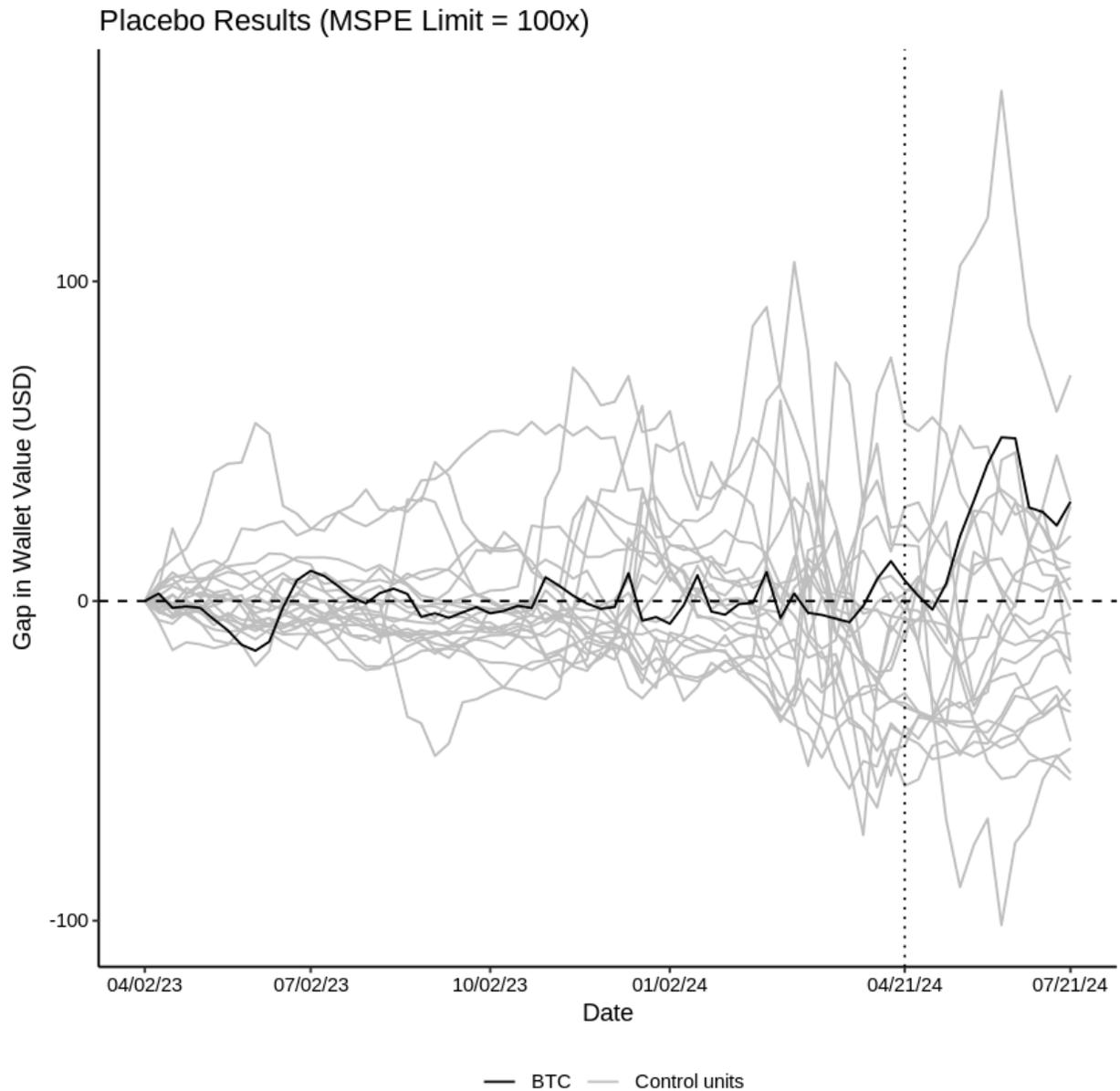

**Figure 31**. This plot shows the difference between the observed unit and synthetic controls for the treated and control units, in black and grey, respectively. Two cryptocurrencies with pre-Halving Mean Squared Prediction Error (MSPE) 100 times higher than Bitcoin's are discarded. Wallet value gaps for Bitcoin and 21 remaining placebo gaps for control cryptocurrencies are shown. The vertical dotted line shows the time of the 2024 Bitcoin halving.

Placebo study with no MSPE limit:

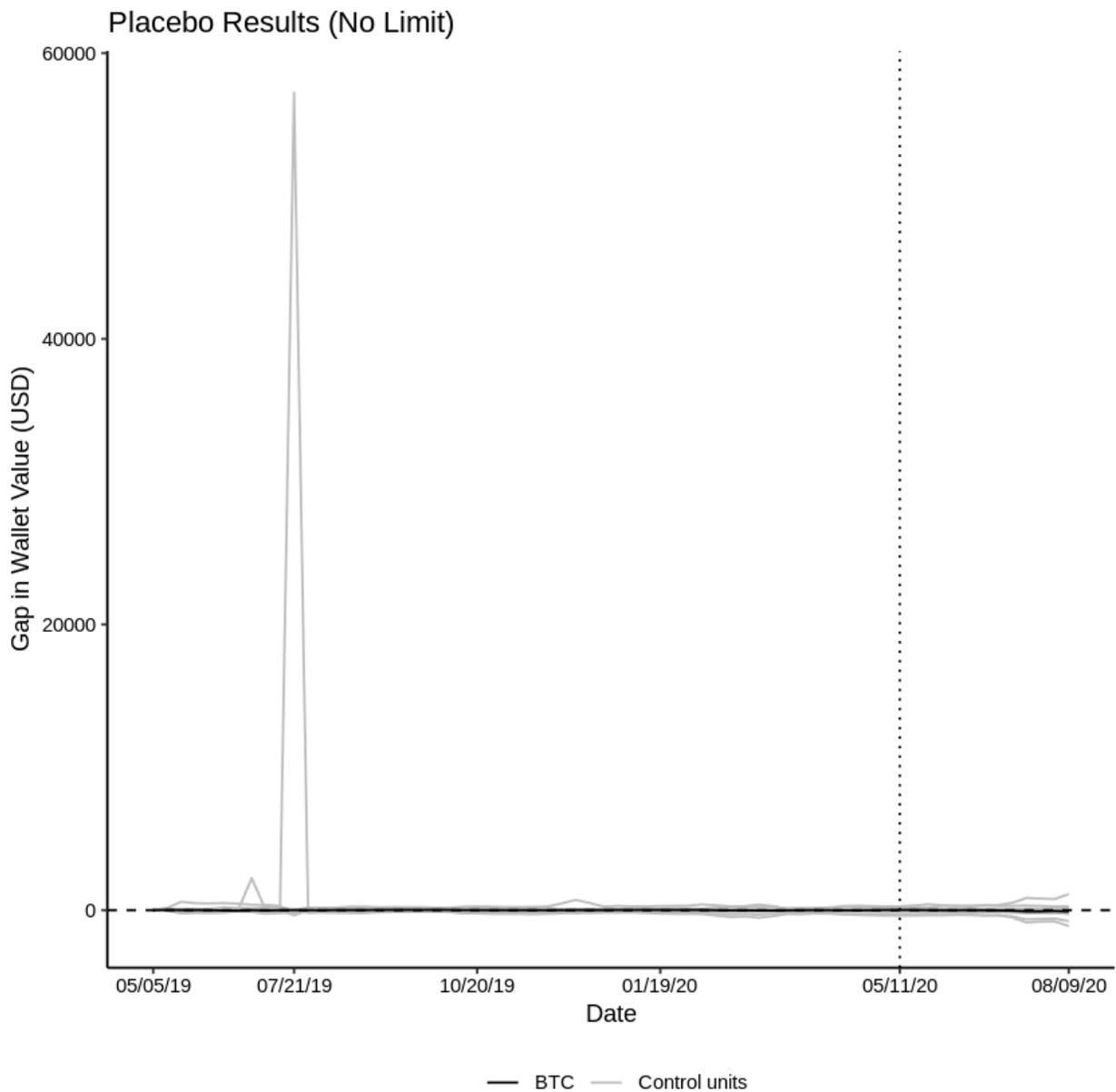

**Figure 32**. This plot shows the difference between observed units and synthetic controls for the treated and control units, in black and grey, respectively. No cryptocurrencies are discarded. Wallet value gaps for Bitcoin and the placebo gaps for 19 control cryptocurrencies are shown. The vertical dotted line shows the time of the 2020 Bitcoin halving.

Placebo study with 100x MSPE limit:

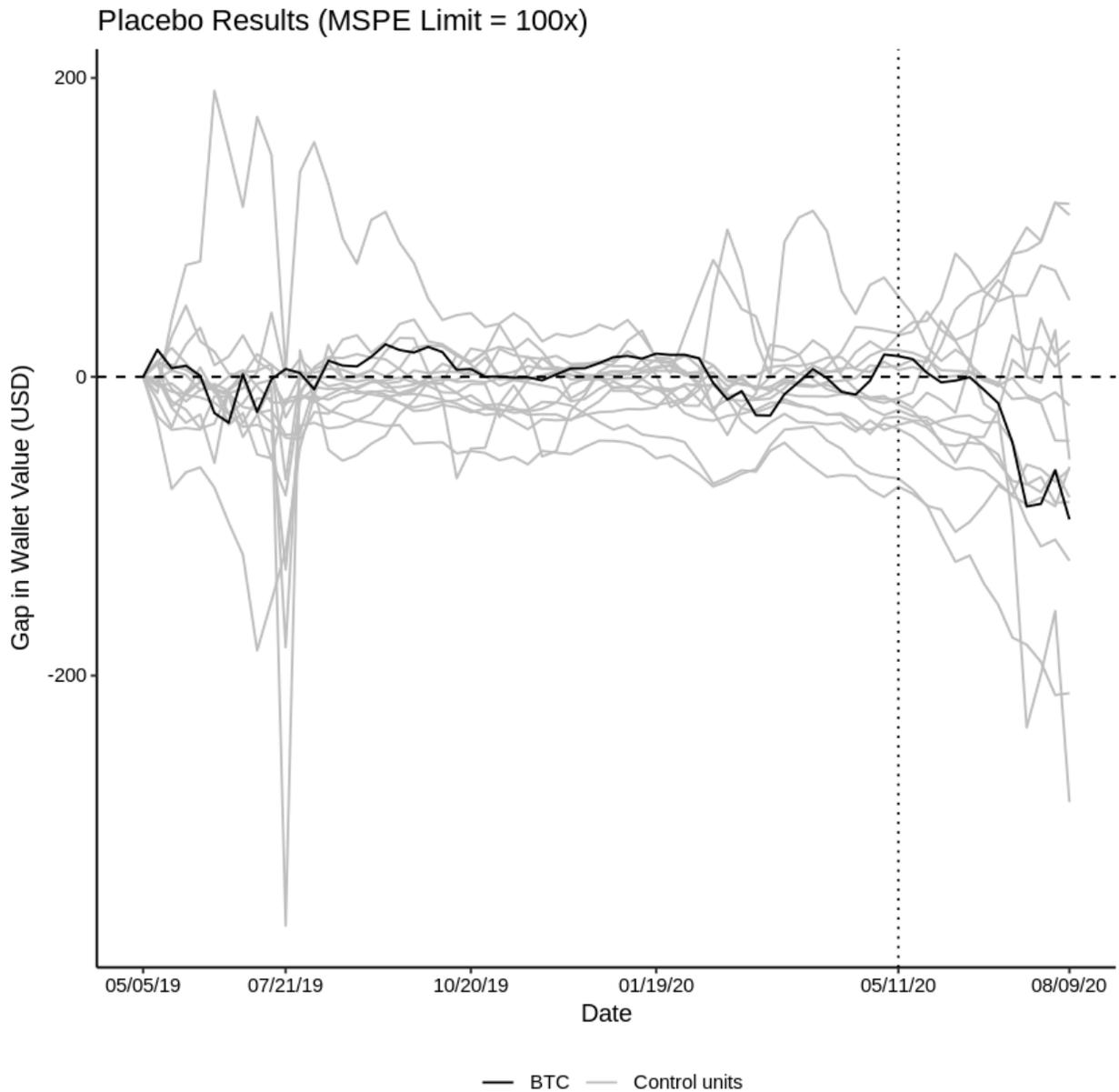

**Figure 33**. This plot shows the difference between the observed unit and synthetic controls for the treated and control units, in black and grey, respectively. Four cryptocurrencies with pre-Halving Mean Squared Prediction Error (MSPE) 100 times higher than Bitcoin's are discarded. Wallet value gaps for Bitcoin and 15 remaining placebo gaps for control cryptocurrencies are shown. The vertical dotted line shows the time of the 2020 Bitcoin halving.